\pdfoutput=1

\documentclass[twocolumn]{aastex62}
\usepackage{amsmath}
\usepackage{graphicx}
\usepackage{mathrsfs}
\usepackage{color}
\usepackage{url}
\usepackage{CJKutf8}
\usepackage{ulem}

\received{2019}
\revised{2019}
\accepted{\today}
\submitjournal{AJ}

\shorttitle{C/2010 U3}
\shortauthors{Hui et al. 2019}

\begin{document}

\title{
C/2010 U3 (Boattini): \\
A Bizarre Comet Active at Record Heliocentric Distance
}

\correspondingauthor{Man-To Hui}
\email{pachacoti@ucla.edu}

\author{
\begin{CJK}{UTF8}{bsmi}
Man-To Hui (許文韜)
\end{CJK}
}
\affiliation{Department of Earth, Planetary and Space Sciences, UCLA,
595 Charles Young Drive East,
Los Angeles, CA 90095-1567, USA}
\nocollaboration

\author{Davide Farnocchia}
\affiliation{Jet Propulsion Laboratory, California Institute of Technology,
4800 Oak Grove Drive, Pasadena, CA 91109, USA}
\nocollaboration

\author{Marco Micheli}
\affiliation{ESA NEO Coordination Centre, 
Largo Galileo Galilei, 1, I-00044 Frascati (RM), Italy}
\affiliation{INAF--Osservatorio Astronomico di Roma, 
Via Frascati, 33, I-00040 Monte Porzio Catone (RM), Italy}
\nocollaboration

\begin{abstract}

We present a photometric and dynamical study of comet C/2010 U3 (Boattini), which was seen active in prediscovery data as early as 2005 November at a new inbound record heliocentric distance $r_{\rm H} = 25.8$ au. Two outburst events around 2009 and 2017 were observed. The coma and tail of the comet consist of dust grains of $\sim$10 \micron in radius, ejected protractedly at speeds $\lesssim$50 m s$^{-1}$ near the subsolar point, and are subjected to the Lorentz force, solar gravitation and radiation pressure force altogether. The prolonged activity indicates that sublimation of supervolatiles (e.g., CO, CO$_2$) is at play, causing a net mass-loss rate $\gtrsim$1 kg s$^{-1}$. To sustain the mass loss, the nucleus radius has to be $\gtrsim$0.1 km. The color of the cometary dust, similar to other long-period comets, is redder than the solar colors, but we also observed potential color variations when the comet was at $10 < r_{\rm H} < 15$ au, concurrent with the onset of crystallisation of amorphous water ice, if at all. Using publicly available and our refined astrometric measurements, we estimated the precise trajectory of the comet and propagated it backward to its previous perihelion. We found that the comet has visited the planetary region $1.96 \pm 0.04$ Myr ago, with barycentric perihelion distance $q = 8.364 \pm 0.004$ au. Thus, C/2010 U3 (Boattini) is almost certainly a dynamically old comet from the Oort cloud, and the observed activity cannot be caused by retained heat from the previous apparition.

\end{abstract}

\keywords{
comets: general --- comets: individual (C/2010 U3) --- methods: data analysis
}

\section{Introduction}

In the past, very few comets were detected and observed to be active at large heliocentric distances; the majority of them show activity within heliocentric distance $r_{\rm H} \approx 5$ au, which is consistent with activity driven by sublimation of water ice. In recent years, an increasing number of all-sky surveys equipped with advancing wide-field cameras, high-speed computers, and mature automatic detection pipelines have led to a rapid growth in discoveries of distant comets, whose activity at great distances cannot be explained by sublimation of water ice because of the low temperatures, but requires different mechanisms such as crystallisation and annealing of amorphous water ice \citep{1992A&A...258L...9P,2009Icar..201..719M}, and sublimation of supervolatiles \citep{2012ApJ...758...29A}. Noteworthily, long-period comet C/2017 K2 (PANSTARRS; hereafter ``K2") was discovered at inbound $r_{\rm H} = 15.9$ au, and later identified in archival serendipitous data at a record distance $r_{\rm H} = 23.7$ au \citep{2017ApJ...847L..19J, 2017ApJ...849L...8M, 2018AJ....155...25H}.

In this work, we present a photometric and dynamical study of a similar long-period comet C/2010 U3 (Boattini; hereafter ``U3"), which was identified by us to be active at even greater heliocentric distances pre-perihelion ($24.6 \le r_{\rm H} \le 25.8$ au) in serendipitous archival data from the 3.6 m Canada-France-Hawaii Telescope (CFHT). The comet was first discovered by A. Boattini in images taken with the Mt. Lemmon 1.5 m reflector on 2010 October 31, with a tiny coma at magnitude $\sim$19, at an inbound heliocentric distance of $r_{\rm H} = 18.4$ au \citep{2010CBET.2535....1B}. To date this remains the most distant discovery of an active comet \citep{2017ApJ...849L...8M}. Similar to orbits of other long-period comets, the orbit of U3 is highly eccentric and inclined, with a perihelion passage in early 2019 at perihelion distance $q = 8.5$ au. Thus, it is also one of the known comets with the largest $q$.\footnote{According to the JPL Small-Body Database Search Engine, there are only 15 known comets with even larger $q$ than that of U3 (retrieved on 2019 January 15).}

The scientific importance of U3 is that, first, it provides us with a precious opportunity to constrain and understand the development of the cometary activity starting from an unprecedentedly seen region, and second, it forms a direct comparison to K2, which helps reveal commonality and diversity of ultra-distant comets. We structure the paper as follows. Section \ref{sec_obs} details the observations we used. Results and discussions are held in Sections \ref{sec_rslt} and \ref{sec_disc}, respectively, while Section \ref{sec_sum} presents a summary.

\begin{deluxetable*}{lcccccccccc}
\tablecaption{Observing Geometry of Comet C/2010 U3 (Boattini)
\label{tab_vgeo}}
\tablewidth{0pt}
\tablehead{ 
\colhead{Date (UT)} & \colhead{Telescope\tablenotemark{a}} & \colhead{Filter} & \colhead{$t_{\rm exp}$ (s)\tablenotemark{b}} & \colhead{$r_{\rm H}$ (au)\tablenotemark{c}}  & 
\colhead{$\it \Delta$ (au)\tablenotemark{d}} & \colhead{$\alpha$ (\degr)\tablenotemark{e}} & 
\colhead{$\varepsilon$ (\degr)\tablenotemark{f}} &
\colhead{$\theta_{-\odot}$ (\degr)\tablenotemark{g}} &
\colhead{$\theta_{-{\bf v}}$ (\degr)\tablenotemark{h}} &
\colhead{$\psi$ (\degr)\tablenotemark{i}}
}
\startdata
2005 Nov 05 & CFHT & {\it i}' & 615 & 25.751 & 24.835 & 0.9 & 157.1 & 16.1 & 196.3 & 0.0\\
\hline
\vspace{-0.2cm} & & {\it r}' & 576 &  &  &  & &&& \\ \vspace{-0.2cm} 
2006 Aug 18 & CFHT & & & 24.620 & 24.241 & 2.2 & 110.9 & 257.2 & 198.7 & 1.9\\
 & & {\it z}' & 560 & &  &  & &&& \\
 \hline
2006 Aug 29\tablenotemark{\dag} & CFHT & {\it z}' & 560 & 24.577 & 24.041 & 2.0 & 121.0 & 261.1 & 198.5 & 1.8\\
\hline
2006 Aug 30 & CFHT & {\it g}' & 420 & 24.572 & 24.022 & 2.0 & 122.1 & 261.5 & 198.5 & 1.8\\
\hline
2009 Sep 16 & SDSS & {\it g}', {\it r}', {\it i}' & 54 & 20.062 & 19.350 & 2.1 & 134.0 & 261.0 & 198.7 & 1.9\\
\hline
\vspace{-0.2cm} &  & {\it B} & 180 & &  &  &  &&&\\ \vspace{-0.2cm}
2011 Jan 30 & Keck &  & & 17.981 & 18.006 & 3.1 & 87.0 & 71.4 & 193.2 & -2.6\\
 &  & {\it V}, {\it R}, {\it I} & 160 &  &  &  &  &&&\\
 \hline
\vspace{-0.2cm} &  & {\it B} & 340 &  &  &  &  &&&\\ \vspace{-0.2cm}
2012 Oct 13 & Keck & & & 15.380 & 14.483 & 1.7 & 153.3 & 246.6 & 199.0 & 1.3\\
 &  & {\it V}, {\it R} & 300 & &  & &  &&&\\
 \hline
\vspace{-0.2cm} &  & {\it B} & 340 & & &  &  &&&\\ \vspace{-0.2cm}
2012 Oct 14 & Keck & & & 15.375 & 14.470 & 1.6 & 154.4 & 246.2 & 198.9 & 1.2\\
 &  & {\it V}, {\it R} & 300 & &  &  & &&& \\
 \hline
\vspace{-0.2cm} & & {\it B} & 300 & & & & &&&\\ \vspace{-0.2cm}
2016 Dec 09 & WIYN & & & 9.681 & 8.911 & 3.8 & 139.5 & 159.2 & 209.1 & -2.9\\
 & & {\it V}, {\it R}, {\it I} & 240 & & & & &&&\\
\hline
2017 Mar 25 & WIYN & {\it B}, {\it V}, {\it R} & 240 & 9.399 & 9.665 & 5.8 & 71.7 & 59.5 & 206.5 & -3.0\\
\hline
2017 Nov 14 & WIYN & {\it V}, {\it R} & 240 & 8.886 & 8.397 & 5.7 & 116.8 & 241.9 & 250.5 & -0.9\\
\hline
2017 Nov 17 & WIYN & {\it B}, {\it V}, {\it R}, {\it I} & 240 & 8.881 & 8.373 & 5.6 & 118.1 & 238.7 & 250.2 & -1.2\\
\hline
2018 Dec 09 & WIYN & {\it B}, {\it V}, {\it R}, {\it I} & 240 & 8.458 & 8.108 & 6.4 & 107.6 & 288.3 & 318.6 & -3.2\\
\hline
2018 Dec 12 & WIYN & {\it B}, {\it V}, {\it R} & 240 & 8.457 & 8.087 & 6.3 & 108.9 & 285.8 & 319.1 & -3.5\\
\hline
2018 Dec 13 & WIYN & {\it B}, {\it V}, {\it R}, {\it I} & 240 & 8.457 & 8.081 & 6.3 & 109.3 & 284.9 & 319.3 & -3.6\\
\enddata
\tablenotetext{a}{CFHT = 3.6 m Canada-France-Hawaii Telescope, Keck = Keck I 10 m telescope, SDSS = Sloan Digital Sky Survey 2.5 m telescope, WIYN = WIYN 0.9 m telescope.}
\tablenotetext{b}{Individual exposure time.}
\tablenotetext{c}{Heliocentric distance.}
\tablenotetext{d}{Topocentric distance.}
\tablenotetext{e}{Phase angle (Sun-comet-observer).}
\tablenotetext{f}{Solar elongation (Sun-observer-comet).}
\tablenotetext{g}{Position angle of projected antisolar direction.}
\tablenotetext{h}{Position angle of projected negative heliocentric velocity of the comet.}
\tablenotetext{i}{Observer to comet's orbital plane angle with vertex at the comet. Negative values indicate observer below the orbital plane of the comet.}
\tablenotetext{\dag}{Only a single image was taken when the comet happened to be apparently overlapped with a background source. So it is obsolete for analysis.}

\end{deluxetable*}

\begin{deluxetable*}{lccccccc}
\tablecaption{Photometry of Comet C/2010 U3 (Boattini)
\label{tab_phot}}
\tablewidth{0pt}
\tablehead{ 
\colhead{Date (UT)} & \colhead{Telescope\tablenotemark{a}} & \colhead{Number\tablenotemark{b}} & \colhead{$\lambda$\tablenotemark{c}} & \colhead{$m_{\lambda} \left(r_{\rm H}, {\it \Delta}, \alpha \right)$\tablenotemark{d}} & \colhead{$m_{\lambda} \left(1,1,0\right)$\tablenotemark{e}}  & 
\colhead{Color} & \colhead{$C_{\rm e}$ ($10^{4}$ km$^{2}$)\tablenotemark{f}}
}
\startdata
2005 Nov 05 & CFHT & 2 & {\it i}' & $22.48 \pm 0.06$ & $8.41 \pm 0.06$ & N/A & $1.25 \pm 0.10$ \\ \hline
2006 Aug 18\tablenotemark{\dag} & CFHT & 3 & {\it r}' & $22.37 \pm 0.05$ & $8.39 \pm 0.05$ & N/A & $1.22 \pm 0.06$ \\ \hline
2006 Aug 30 & CFHT & 4 & {\it g}' & $22.98 \pm 0.09$ & $9.04 \pm 0.09$ & N/A & $1.22 \pm 0.06$ \\ \hline
\vspace{-0.15cm} 2009 Sep 16 &  SDSS & 1 & {\it g}' & $21.59 \pm 0.08$ & $8.55 \pm 0.08$ & {\it g}' $-$ {\it r}' = $+0.61 \pm 0.11$ & $1.88 \pm 0.10$ \\
\vspace{-0.15cm} & & 1 & {\it r}' & $20.97 \pm 0.08$ & $7.94 \pm 0.08$ & {\it r}' $-$ {\it i}' = $-0.04 \pm 0.14$ & \\
 &  & 1& {\it i}' & $21.02 \pm 0.11$ & $7.98 \pm 0.11$ & {\it g}' $-$ {\it i}' = $+0.57 \pm 0.14$ & \\ \hline
\vspace{-0.15cm}2011 Jan 30 & Keck & 6 & {\it B} & $21.72 \pm 0.03$ & $9.03 \pm 0.03$ & {\it B} $-$ {\it V} = $+0.77 \pm 0.04$ & $1.76 \pm 0.03$ \\
\vspace{-0.15cm} &  & 2 & {\it V} & $20.95 \pm 0.02$ & $8.26 \pm 0.02$ & {\it B} $-$ {\it R} = $+1.21 \pm 0.04$ & \\\vspace{-0.15cm}
 &  & 2 & {\it R} & $20.51 \pm 0.01$ & $7.82 \pm 0.01$ & {\it V} $-$ {\it R} = $+0.44 \pm 0.02$ & \\ 
 &  & 2 & {\it I} & $20.13 \pm 0.02$ & $7.44 \pm 0.02$ & {\it R} $-$ {\it I} = $+0.39 \pm 0.02$ & \\ \hline
\vspace{-0.15cm} 2012 Oct 13 & Keck & 4 & {\it B} & $21.01 \pm 0.05$ & $9.20 \pm 0.05$ & {\it B} $-$ {\it V} = $+0.81 \pm 0.07$& $1.56 \pm 0.06$ \\ 
\vspace{-0.15cm} & & 2 & {\it V} & $20.20 \pm 0.04$ & $8.39 \pm 0.04$ & {\it B} $-$ {\it R} = $+1.30 \pm 0.08$ & \\
 & & 1 & {\it R} & $19.71 \pm 0.06$ & $7.89 \pm 0.06$ & {\it V} $-$ {\it R} = $+0.49 \pm 0.07$ & \\ \hline
\vspace{-0.15cm} 2012 Oct 14\tablenotemark{\ddag}
 & Keck & 2 & {\it B} & $21.07 \pm 0.04$ & $9.26 \pm 0.04$ & {\it B} $-$ {\it V} = $+0.79 \pm 0.06$ & $1.45 \pm 0.06$ \\ \vspace{-0.15cm}
 & & 1 & {\it V} & $20.28 \pm 0.05$ & $8.47 \pm 0.05$ & {\it B} $-$ {\it R} = $+1.34 \pm 0.08$ & \\
 & & 1 & {\it R} & $19.73 \pm 0.06$ & $7.92 \pm 0.06$ & {\it V} $-$ {\it R} = $+0.55 \pm 0.08$ & \\ \hline
 \vspace{-0.15cm} 2016 Dec 09 & WIYN & 6 & {\it B} & $19.29 \pm 0.09$ & $9.45 \pm 0.09$ & {\it B} $-$ {\it V} = $+1.03 \pm 0.09$ & $1.53 \pm 0.02$ \\
 \vspace{-0.15cm} & & 5 & {\it V} & $18.26 \pm 0.02$ & $8.41 \pm 0.02$ & {\it B} $-$ {\it R} = $+1.46 \pm 0.10$ & \\ \vspace{-0.15cm}
  & & 5 & {\it R} & $17.83 \pm 0.03$ & $7.99 \pm 0.03$ & {\it V} $-$ {\it R} = $+0.43 \pm 0.04$ & \\
  & & 5 & {\it I} & $17.28 \pm 0.06$ & $7.44 \pm 0.06$ & {\it R} $-$ {\it I} = $+0.55 \pm 0.07$ & \\ \hline
\vspace{-0.15cm} 2017 Mar 25 & WIYN & 4 & {\it B} & $19.32 \pm 0.03$ & $9.29 \pm 0.03$ & {\it B} $-$ {\it V} = $+0.98 \pm 0.03$ & $1.69 \pm 0.03$ \\ 
\vspace{-0.15cm}
  & & 4 & {\it V} & $18.34 \pm 0.02$ & $8.30 \pm 0.02$ & {\it B} $-$ {\it R} = $+1.43 \pm 0.05$ & \\
  & & 4 & {\it R} & $17.89 \pm 0.04$ & $7.86 \pm 0.04$ & {\it V} $-$ {\it R} = $+0.45 \pm 0.05$ & \\ \hline
\vspace{-0.15cm} 2017 Nov 14 & WIYN & 4 & {\it V} & $18.01 \pm 0.04$ & $8.40 \pm 0.04$ & {\it V} $-$ {\it R} = $+0.48 \pm 0.06$ & $1.55 \pm 0.06$ \\ 
  & & 4 & {\it R} & $17.53 \pm 0.04$ & $7.92 \pm 0.04$ & & \\ \hline
\vspace{-0.15cm} 2017 Nov 17 & WIYN & 4 & {\it B} & $18.90 \pm 0.05$ & $9.30 \pm 0.05$ & {\it B} $-$ {\it V} = $+0.94 \pm 0.05$ & $1.60 \pm 0.02$\\ 
\vspace{-0.15cm} 
  & & 4 & {\it V} & $17.96 \pm 0.01$ & $8.36 \pm 0.01$ & {\it B} $-$ {\it R} = $+1.46 \pm 0.06$ & \\ \vspace{-0.15cm}
  & & 4 & {\it R} & $17.44 \pm 0.04$ & $7.84 \pm 0.04$ & {\it V} $-$ {\it R} = $+0.52 \pm 0.04$ & \\
  & & 4 & {\it I} & $16.91 \pm 0.04$ & $7.31 \pm 0.04$ & {\it R} $-$ {\it I} = $+0.53 \pm 0.06$ & \\ \hline
\vspace{-0.15cm} 2018 Dec 09 & WIYN & 4 & {\it B} & $18.96 \pm 0.04$ & $9.51 \pm 0.04$ & {\it B} $-$ {\it V} = $+0.83 \pm 0.05$ & $1.19 \pm 0.03$\\ 
\vspace{-0.15cm} 
  & & 4 & {\it V} & $18.13 \pm 0.03$ & $8.68 \pm 0.03$ & {\it B} $-$ {\it R} = $+1.34 \pm 0.05$ & \\ \vspace{-0.15cm}
  & & 4 & {\it R} & $17.62 \pm 0.03$ & $8.17 \pm 0.03$ & {\it V} $-$ {\it R} = $+0.51 \pm 0.04$ & \\
  & & 4 & {\it I} & $17.18 \pm 0.04$ & $7.73 \pm 0.04$ & {\it R} $-$ {\it I} = $+0.44 \pm 0.05$ & \\ \hline
\vspace{-0.15cm} 2018 Dec 12 & WIYN & 6 & {\it B} & $19.03 \pm 0.05$ & $9.59 \pm 0.05$ & {\it B} $-$ {\it V} = $+0.90 \pm 0.06$ & $1.18 \pm 0.02$\\ 
\vspace{-0.15cm} 
  & & 6 & {\it V} & $18.13 \pm 0.02$ & $8.69 \pm 0.02$ & {\it B} $-$ {\it R} = $+1.39 \pm 0.05$ & \\
  & & 6 & {\it R} & $17.64 \pm 0.01$ & $8.20 \pm 0.01$ & {\it V} $-$ {\it R} = $+0.49 \pm 0.03$ & \\ \hline
\vspace{-0.15cm} 2018 Dec 13 & WIYN & 6 & {\it B} & $18.90 \pm 0.09$ & $9.46 \pm 0.09$ & {\it B} $-$ {\it V} = $+0.83 \pm 0.11$ & $1.25 \pm 0.06$\\ 
\vspace{-0.15cm} 
  & & 6 & {\it V} & $18.07 \pm 0.05$ & $8.63 \pm 0.05$ & {\it B} $-$ {\it R} = $+1.34 \pm 0.09$ & \\ \vspace{-0.15cm}
  & & 6 & {\it R} & $17.55 \pm 0.01$ & $8.11 \pm 0.01$ & {\it V} $-$ {\it R} = $+0.52 \pm 0.05$ & \\
  & & 6 & {\it I} & $17.12 \pm 0.08$ & $7.68 \pm 0.08$ & {\it R} $-$ {\it I} = $+0.44 \pm 0.08$ & \\
\enddata
\tablenotetext{a}{CFHT = 3.6 m Canada-France-Hawaii Telescope, Keck = Keck I 10 m telescope, SDSS = Sloan Digital Sky Survey 2.5 m telescope, WIYN = WIYN 0.9 m telescope.}
\tablenotetext{b}{Number of used exposures.}
\tablenotetext{c}{Reduction bandpass.}
\tablenotetext{d}{Apparent magnitude and associated uncertainty in the corresponding reduction bandpass.}
\tablenotetext{e}{Absolute magnitude.}
\tablenotetext{f}{Effective scattering cross-section in $10^4$ km$^{2}$, estimated from Equation (\ref{eq_XS}) assuming geometric albedo $p_{V} = 0.04$.}
\tablenotetext{\dag}{Unfortunately, the comet entirely blended with a background galaxy in all the {\it z}'-band images from this date so no photometry was performed.}
\tablenotetext{\ddag}{The comet was close to an overexposed field star, whose halo was removed using azimuthal median subtraction centering on the star before photometry was performed.}

\tablecomments{
All of the photometry data were measured using a circular aperture of $\varrho = 3.5 \times 10^4$ km radius. Weighted mean values of apparent magnitude of the comet are reported for multiple-exposure observations, and the uncertainties are their standard deviation from repeated measurements. In cases where there is only a single useful exposure from a given night, the uncertainty is determined by the SNR of the comet, CCD gain and readout noise values. 
}
\end{deluxetable*}

\section{Observations}
\label{sec_obs}

We collected images of U3 taken in 2011-2012 from the Keck I 10 m telescope, and conducted observations of it with the WIYN\footnote{The WIYN Observatory is a joint facility of the University of Wisconsin-Madison, Indiana University, the National Optical Astronomy Observatory and the University of Missouri.} 0.9 m telescope in 2016-2018. By using the Solar System Object Image Search service \citep{2012PASP..124..579G} at the Canadian Astronomy Data Centre, we obtained serendipitous prediscovery archival data from the CFHT and the Sloan Digital Sky Survey (SDSS) telescope of the comet. The observing geometry and conditions of U3 are summarized in Table \ref{tab_vgeo}.

\subsection{Canada-France-Hawaii Telescope}
\label{sec_cfht}

The 3.6 m f/4.1 CFHT is located atop Mauna Kea, Hawai`i. The prediscovery {\it g'}-, {\it r'}-, {\it i'}- and {\it z'}-band  images containing the comet were taken from 2005 November 05, and 2006 August 18 and 29-30 by the MegaCam prime focus imager. The exposure duration varies for different sets of filters (see Table \ref{tab_vgeo}). Each image consists of 36 subfields having a common field-of-view (FOV) of $6\farcm4 \times 12\farcm8$ and an angular resolution of 0\farcs187 pixel$^{-1}$. Although the telescope was tracked at the sidereal rate, thanks to the slow apparent motion of the comet ($\lesssim$5\farcs4 hr$^{-1}$), images of the comet remain untrailed during exposures in all of the CFHT data only at subpixel levels. The full width at half maximum (FWHM) of the field stars in the images varied between $\sim$0\farcs6 and 1\farcs1.

\subsection{Keck I Telescope}
\label{sec_keck}

Optical images of U3 were obtained through the Keck I 10 m telescope on the Mauna Kea, Hawai`i with the Low-Resolution Imaging Spectrometer (LRIS) camera \citep{1995PASP..107..375O} from 2011 January 30, 2012 October 13 and 14. The LRIS camera has independent blue and red channels separated by a dichroic beam splitter. The ``460" dichroic, which has 50\% transmission at wavelength 4875 \AA, was exploited. On the blue side, a broadband {\it B} filter with effective wavelength $\lambda_{\rm eff} = 4370$ \AA~and FWHM $\Delta \lambda = 878$ \AA~was used. On the red side, {\it V}-band ($\lambda_{\rm eff} = 5473$ \AA, FWHM $\Delta \lambda = 948$ \AA) and {\it R}-band ($\lambda_{\rm eff} = 6417$ \AA, FWHM $\Delta \lambda = 1185$ \AA) filters were used for all the three nights. For the 2011 observation, images of the comet through an {\it I}-band filter ($\lambda_{\rm eff} = 7599$ \AA, FWHM $\Delta \lambda = 1225$ \AA) were also taken. All of the observations exploited an atmospheric dispersion compensator to correct for differential refraction, and the telescope was tracked on the apparent motion of the comet nonsidereally with autoguiding on fixed stars. Exposure durations for {\it B}-band images were longer than for other images from the same nights (see Table \ref{tab_vgeo}). The images, calibrated with bias subtraction and flat fielding using images of a diffusely illuminated patch on the inside of the Keck dome, have a pixel scale of 0\farcs135 pixel$^{-1}$, and a useful FOV of $\sim$$7\farcm5 \times 6\farcm0$. The FWHM values of the field stars varied between $\sim$0\farcs6-1\farcs4 FWHM in the images.

\subsection{Sloan Digital Sky Survey Telescope}
\label{sec_sdss}

\begin{figure*}
\epsscale{1.0}
\begin{center}
\plotone{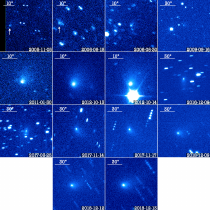}
\caption{
Composite images of comet C/2010 U3 (Boattini), with dates in UT and scale bars labelled in each panel. The Keck and WIYN images are all in {\it R} band. Images from 2006 August 18 and 2009 September 16 are coaddition from data in different bands. Equatorial north is up and east is left. We do not show the antisolar or the projected heliocentric velocity directions to avoid cluttering the plots, but list them in Table \ref{tab_vgeo}. The comet was close to the edge of the images from the first two nights.
\label{fig_U3}
} 
\end{center} 
\end{figure*}

Three prediscovery {\it g}'-, {\it r}'- and {\it i}'-band images taken by the SDSS 2.5 m f/5 Ritchey-Chr{\'e}tien telescope at Apache Point Observatory, New Mexico, on 2009 September 16, in which comet U3 was visible, were found. We could not find the comet in the {\it u}'- and {\it z}'-band images due to its faintness. The CCD FOV is $13\farcm5 \times 9\farcm0$, while the pixel scale is 0\farcs396 pixel$^{-1}$. A common exposure ($t_{\rm exp} = 54$ s) was exploited for the all of the images, during which the comet was not trailed because of its slow apparent motion. In fact, hardly can we discern the displacement of the comet across the images. However, the identification is utterly unambiguous by checking deeper images which show no background sources of similar brightness whatsoever at the position. The average FWHM of the field stars varies little, from $\sim$1\farcs0 in the {\it i}'-band image to 1\farcs1 in the {\it g}'-band one, whereas the comet obviously appeared nonstellar (Figure \ref{fig_U3}).

\subsection{WIYN 0.9m Observatory}
\label{sec_KP}

We obtained {\it B}-, {\it V}-, {\it R}-, and {\it I}-band images of U3 from the Half Degree Imager (HDI) attached to the WIYN 0.9 m f/7.5 telescope at the Kitt Peak National Observatory, Arizona, from 2016 to 2018.\footnote{Detailed information of the filters can be found at \url{https://www.noao.edu/0.9m/observe/s2kb.html}.} The HDI has an image dimension of $4096 \times 4096$ pixels, while the FOV is $\sim0\fdg49 \times 0\fdg49$, with an angular resolution 0\farcs425 pixel$^{-1}$. Exposure durations of the images were all 240 s, except the {\it B}-band images from 2016 December 09, for which an exposure time of 300 s was used. The telescope was tracked at a nonsidereal rate according to the apparent motion of U3 during our observing runs, such that while the comet remained unblurred, the field background sources were slightly trailed, by a few pixels at most. Field stars in the images have typical FWHM values varying between $\sim$1\farcs3 and 1\farcs9. We calibrated the images with bias frames and corresponding flat-field frames using a diffusely illuminated spot on the inside of the observatory dome.

\section{Results}
\label{sec_rslt}

\subsection{Photometry}
\label{subsec_phot}

We performed aperture photometry of U3 in the images from the aforementioned observatories in Section \ref{sec_obs}. In order to eliminate potential biases from the aperture effect, a spatially fixed aperture radius of $\varrho = 3.5 \times 10^4$ km was used, which is large enough (always more than twice the average FWHM values of field stars) to avoid shape distortion by the point-spread function, fluctuation from seeing and encompass the majority of the flux of U3, while the signal-to-noise ratio (SNR) is not too low. It also helps avoid field stars falling into the aperture in most cases. The sky background was determined from an adjacent annulus extending from twice the photometric aperture radius and the same length in width. We handled the Keck images from 2012 October 14 specially, as the comet was in a halo of an overexposed star only $\sim$7\arcsec~away. Taking advantage of the halo being circularly symmetric around the star, we measured the flux of the star in a series of concentric rings, and obtained the azimuthally median brightness, which was subsequently subtracted from the images, leaving a nicely flattened and clean background without any visible artefact around the comet.

For photometry of stars we chose the photometric aperture radius to be roughly twice the average FWHM values of field stars, while the sky background was measured in an adjacent annulus between $\sim$$3\times$ and $5\times$FWHM from the centroid. For the Keck observations, we calibrated the brightness of U3 using a number of Landolt standard stars \citep{1992AJ....104..340L} at similar airmass. The magnitudes of the stars in the CFHT and WIYN images were calibrated to the SDSS Data Release 12 \citep[DR12;][]{2015ApJS..219...12A} and the Pan-STARRS1 (PS1) Data Release 1 \citep[DR1;][]{2016arXiv161205243F}, respectively. The reason for such a decision is that the filter system at the CFHT is very close to the Sloan system, and the WIYN images are not covered by the SDSS DR12 but the PS1 DR1 catalogue only. For the latter, we had to transform magnitudes from the PS1 photometric system to the Johnson-Bessel one using the relationships derived by \citet{2012ApJ...750...99T}. We estimated the errors in magnitudes of the comet from errors in image zero-points, and the standard deviation of repeated measurements. In cases where there was only one image available from a single night, we calculated the flux errors from Poisson statistics.

For the SDSS images, no photometry of field stars or sky background measurement was conducted, as they had been photometrically calibrated to a unit of ``nanomaggy" with removal of the sky background. Conversion from the fluxes of U3 to the magnitudes was achieved by following steps described in the SDSS document (\url{http://www.sdss3.org/dr8/algorithms/fluxcal.php}). The magnitude uncertainties were determined from Poisson statistics in reconstructed images before the removal of sky background with known values of the CCD gain, readout noise and dark current.

\begin{figure}
\epsscale{1.0}
\begin{center}
\plotone{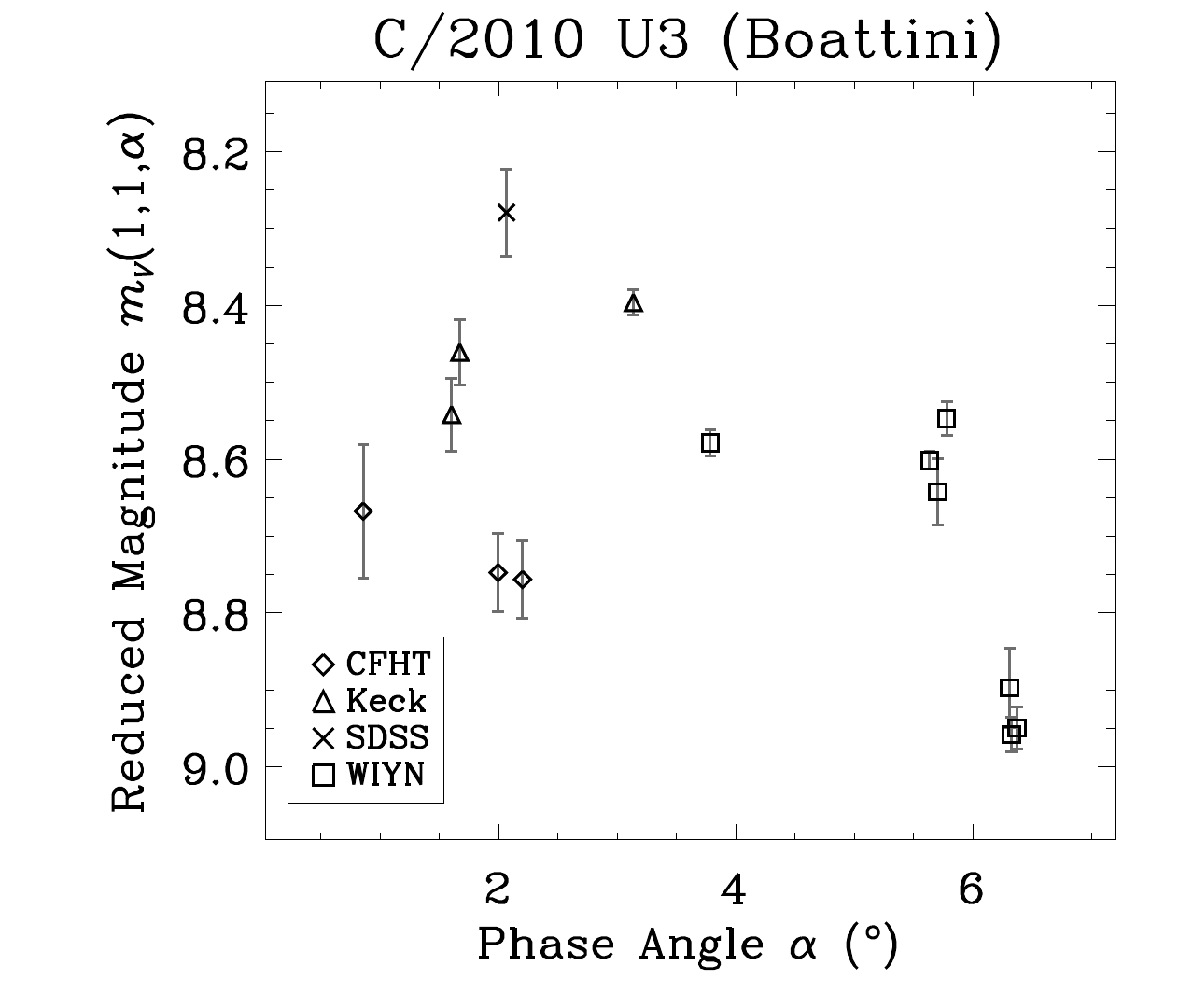}
\caption{
Reduced {\it V}-band magnitude $m_{V} \left(1,1,\alpha \right)$ versus phase angle $\alpha$ for comet C/2010 U3 (Boattini). Note that the scatter of the data points is obviously larger than the magnitude uncertainties, most likely indicating activity variations of the comet, and a backscattering enhancement smaller than the intrinsic variations of the brightness.
\label{fig_U3_phi}
} 
\end{center} 
\end{figure}

\begin{figure*}
\gridline{\fig{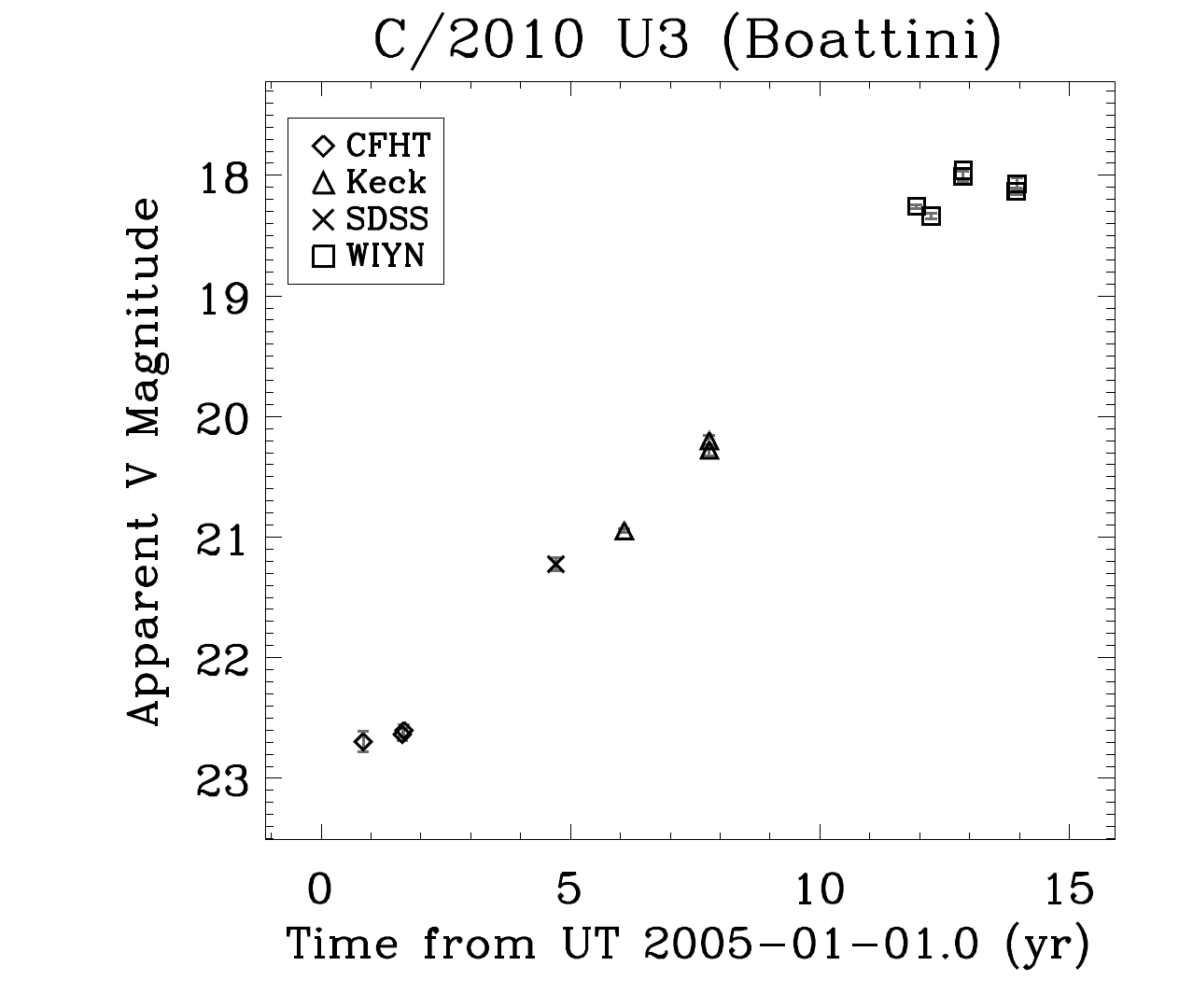}{0.5\textwidth}{(a)}
          \fig{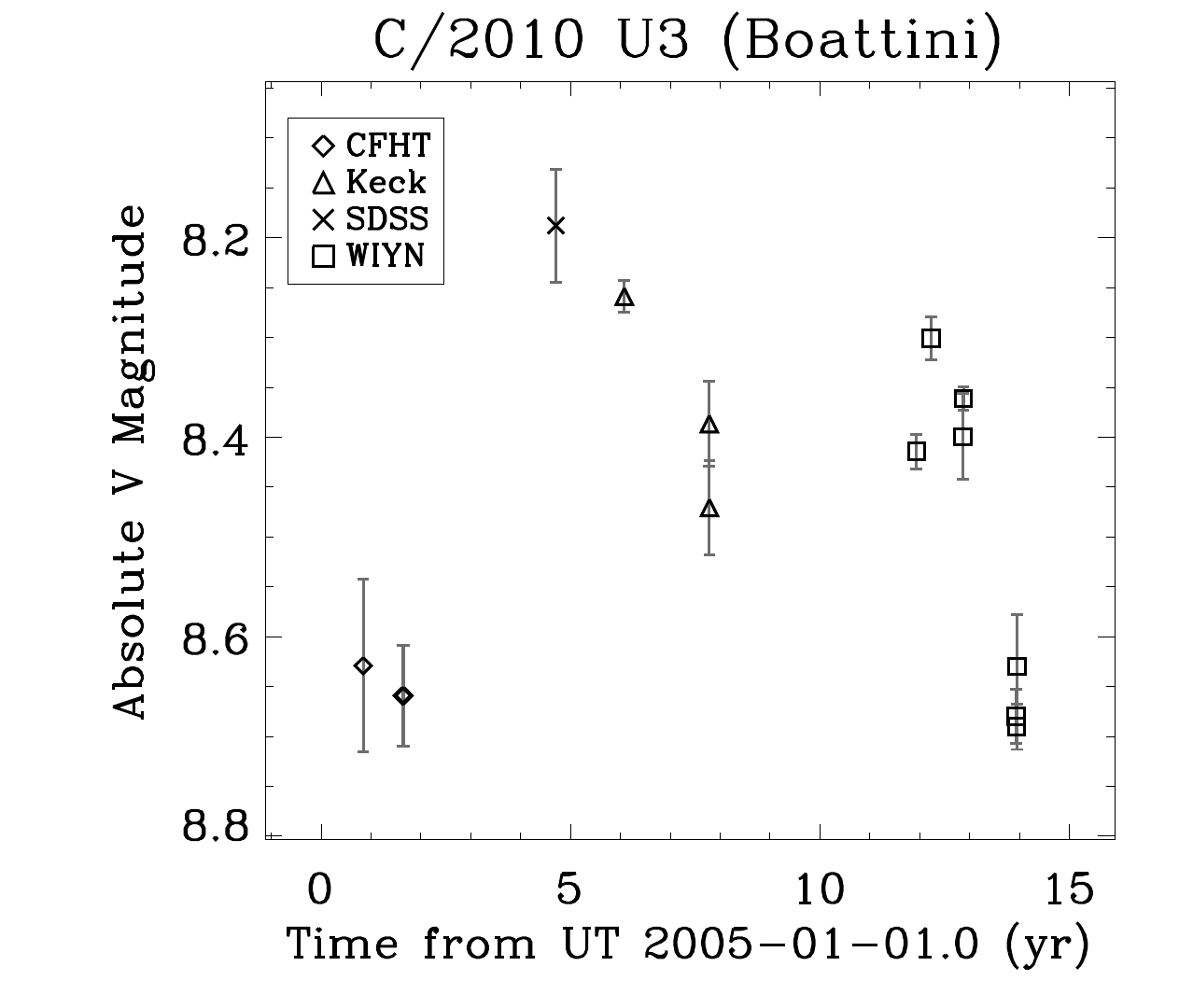}{0.5\textwidth}{(b)}
          }
\caption{
Temporal evolution of $V$-band magnitude of comet C/2010 U3 (Boattini). Point symbols correspond to observatories as indicated in the legend. Apparent magnitude data in panel (a) reduced to $r_{\mathrm{H}} = \mathit{\Delta} = 1$ au and $\alpha = 0$\degr~using Equation (\ref{eq_H}) yields panel (b).
\label{fig_U3_lc}
} 
\end{figure*}

Table \ref{tab_phot} lists the measurements of the apparent magnitudes of U3 in different bandpasses, denoted as $m_{\lambda} \left(r_{\rm H}, {\it \Delta}, \alpha \right)$, where ${\it \Delta}$ is the distance between the comet and the observer, and $\alpha$ is the phase angle. To assess the intrinsic brightness, we corrected for the varying observing geometry and obtained the absolute magnitudes from
\begin{equation}
m_{\lambda} \left(1,1,0\right) = \underbrace{m_{\lambda} \left(r_{\rm H}, {\it \Delta}, \alpha \right) - 5 \log \left(r_{\rm H} {\it \Delta}\right)}_{m_{\lambda} \left(1, 1, \alpha \right)} + 2.5 \log \phi \left(\alpha \right),
\label{eq_H}
\end{equation} 
\noindent where $\phi\left(\alpha\right)$ is the phase function, and $m_{\lambda} \left(1, 1, \alpha \right)$ is the reduced magnitude. Originally we intended to derive $\phi\left(\alpha\right)$ from our data, however, we noticed that the scatter was clearly too large compared to the individual errors in magnitude (Figure \ref{fig_U3_phi}). This is likely due to the fact that the comet exhibited activity variations which obscured the backscattering enhancement. Rather than finding a best-fit phase function from the measurements, we assumed that $\phi \left( \alpha \right)$ could be approximated by the combined Halley-Marcus phase function by \citet{2007ICQ....29...39M} and \citet{2011AJ....141..177S}. Magnitudes from the CFHT and SDSS were transformed from the Sloan system to the Johnson-Cousins one using the observed color of the comet (Table \ref{tab_phot}) and the relationships derived by \citet{Jordi06}. Figure \ref{fig_U3_lc} shows both the apparent and absolute {\it V}-band magnitudes of the comet as a function of time. Although the apparent magnitude brightens almost linearly with time, the intrinsic brightness does not. Intriguingly, Figure \ref{fig_U3_lc}(b) suggests that the comet underwent an outburst event around year 2009 at $r_{\rm H} > 20$ au. Early 2017 witnessed another outburst of the comet, at $\sim$9 au from the Sun.

\begin{deluxetable*}{ll|cc|cc}
\tablecaption{Selected Best-Fit Orbit Solutions of Comet C/2010 U3 (Boattini)
\label{tab_orb}}
\tablewidth{0pt}
\tablehead{
\multicolumn{2}{c|}{Quantity} & 
\multicolumn{2}{c|}{Gravity-Only}  & \multicolumn{2}{c}{CO-Sublimation $A_j$ ($j=1,2,3$)}  \\
& & 
Value & 1$\sigma$ Uncertainty &
Value & 1$\sigma$ Uncertainty
}
\startdata
Perihelion distance (au) & $q$ 
       & 8.4511580 & $1.01\times10^{-5}$
       & 8.4511248 & $2.98\times10^{-5}$ \\
Eccentricity & $e$
       & 1.00327813 & $3.83\times10^{-6}$
       & 1.00326930 & $7.17\times10^{-6}$ \\
Inclination (\degr) & $i$ 
       & 55.47519397 & $9.89\times10^{-6}$
       & 55.4752052 & $1.69\times10^{-5}$ \\
Longitude of ascending node (\degr) & $\Omega$
                 & 43.0448271 & $1.04\times10^{-5}$
                 & 43.0448287 & $1.05\times10^{-5}$ \\
Argument of perihelion (\degr) & $\omega$
                 & 87.980000 & $1.28\times10^{-4}$
                 & 87.980347 & $2.96\times10^{-4}$ \\
Time of perihelion (TDB)\tablenotemark{\dag} & $t_\mathrm{p}$
                  & 2019 Feb 23.98787 & $2.21\times10^{-3}$
                  & 2019 Feb 23.99304 & $4.10\times10^{-3}$ \\
RTN nongravitational parameters (au d$^{-2}$) & $A_1$
                  & N/A & N/A
                  & $+1.49\times10^{-6}$ & $1.01\times10^{-6}$ \\
& $A_2$
                  & N/A & N/A
                  & $+9.18\times10^{-7}$ & $8.26\times10^{-7}$ \\
& $A_3$
                  & N/A & N/A
                  & $-6.02\times10^{-8}$ & $1.19\times10^{-7}$ \\
\enddata
\tablenotetext{\dag}{The corresponding uncertainties are in days.}
\tablecomments{
The epoch of the both best-fit orbits is JD 2456979.5 = TDB 2014 November 18.0, referenced to the J2000 heliocentric ecliptic. We included 739 astrometric observations to obtain the solutions, with $\chi^2 = 240.3$, and normalized RMS 0.404 for the gravity-only solution, and $\chi^2 = 235.9$ and normalized RMS 0.401 for the CO-sublimation $A_j$ ($j=1,2,3$) solution. See Section \ref{subsec_orb} for detailed information.
}
\end{deluxetable*}

\subsection{Orbit Determination}
\label{subsec_orb}

Our astrometric measurements of U3 from the prediscovery CFHT images in 2005-2006, and from the archival Keck data from 2010-2012 showed unacceptably large astrometric errors when combined with the MPC astrometric dataset. The poor fit was an indication of possible problems with the astrometry, possibly due to tailward biases common for comets, or that nongravitational forces were at work and significantly affected the trajectory of U3, which would be uncommon because of the large heliocentric distances.

In order to identify the cause of the large astrometric residuals, we remeasured astrometry from the SDSS and WIYN. We also extended the arc by observing U3 at WIYN in December 2018 together with the prediscovery CFHT data. The size of centroiding aperture we exploited for astrometry roughly depends on FWHM values of the optocentre of the comet, varying between 1.5 and 3.5 pixels in radius. This reduced, high-quality astrometric dataset was used as a calibration to identify problematic observations in the rest of the dataset. As a result of this analysis, we deleted 984 astrometric positions that appeared to have poor internal consistency or to be clearly biased, which is a common problem for comet astrometry. The remaining 739 observations provide complete coverage of the full observation arc from 2005 to 2018 and can satisfactorily be fit with $\chi^2 = 240.3$ by using a gravity-only model, which considers the gravity of the Sun, perturbations from the eight major planets, Moon, Pluto, and the 16 most massive asteroids in the main belt, and the post-Newtonian relativistic corrections \citep{2015aste.book..815F}.

We still investigated the possibility that nongravitational perturbations were affecting the trajectory of U3. Since temperatures at U3 are too low for water ice to sublimate, rather than applying the canonical water-ice sublimation model by \citet{1973AJ.....78..211M}, we adopted the CO-driven outgassing model by \citet{1996MNRAS.283..347Y}. The estimated $A_1$ (radial), $A_2$ (transverse), and $A_3$ (normal) nongravitational parameters were compatible with 0 (SNR $<1.5$ for all the components, see Table \ref{tab_orb}). By adding radial and transverse accelerations, $\chi^2$ lowered by 4.1 relative to the gravity-only fit, which corresponds to a $p$-value of 15\%. Further adding the normal component lowered $\chi^2$ by 4.4 relative to the gravity-only fit, which corresponds to a $p$-value of 25\%. Moreover, we tested predictions for our 2018 December astrometry based on shorter-arc solutions. The solutions that included nongravitational perturbations did not provide better predictions. Therefore, we concluded that the astrometric database contains no clear evidence that nongravitational forces are materially affecting the trajectory of U3. Two of our best-fit orbit solutions of U3 are presented in Table \ref{tab_orb}.

\begin{figure*}
\gridline{\fig{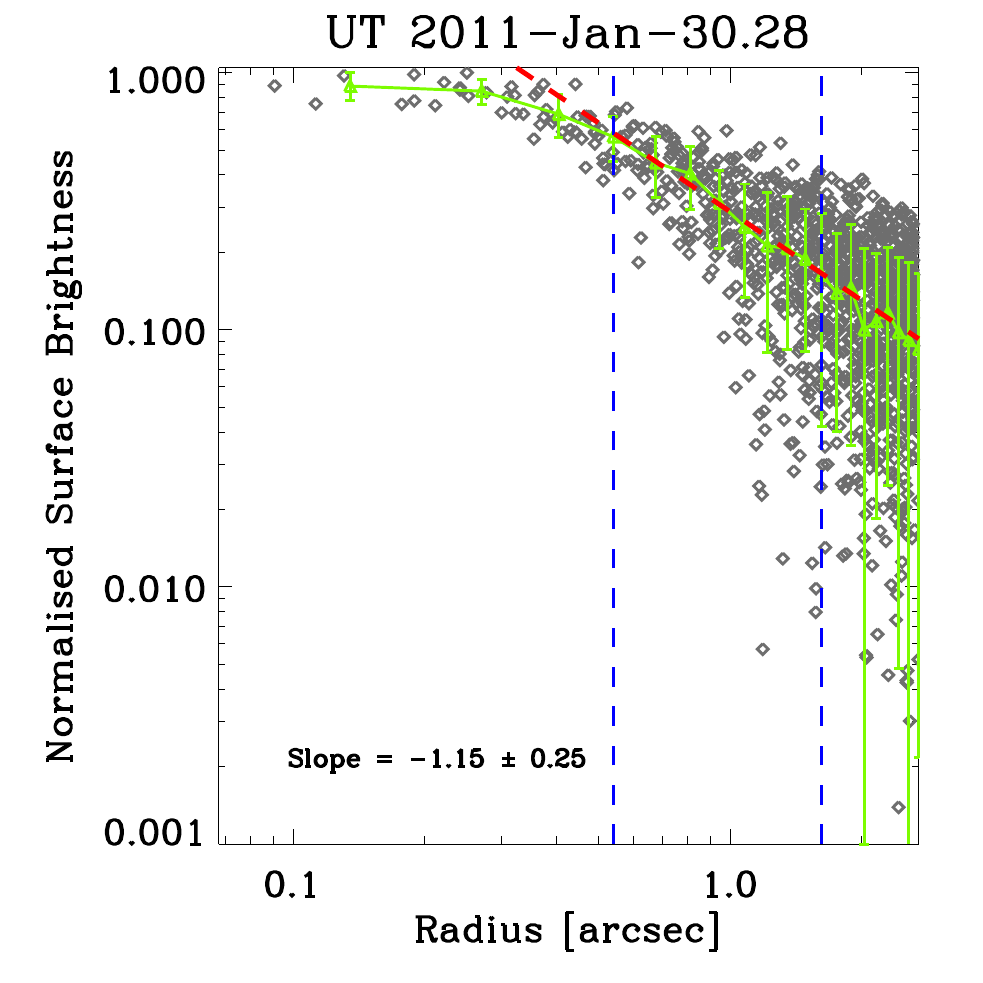}{0.32\textwidth}{(a)}
          \fig{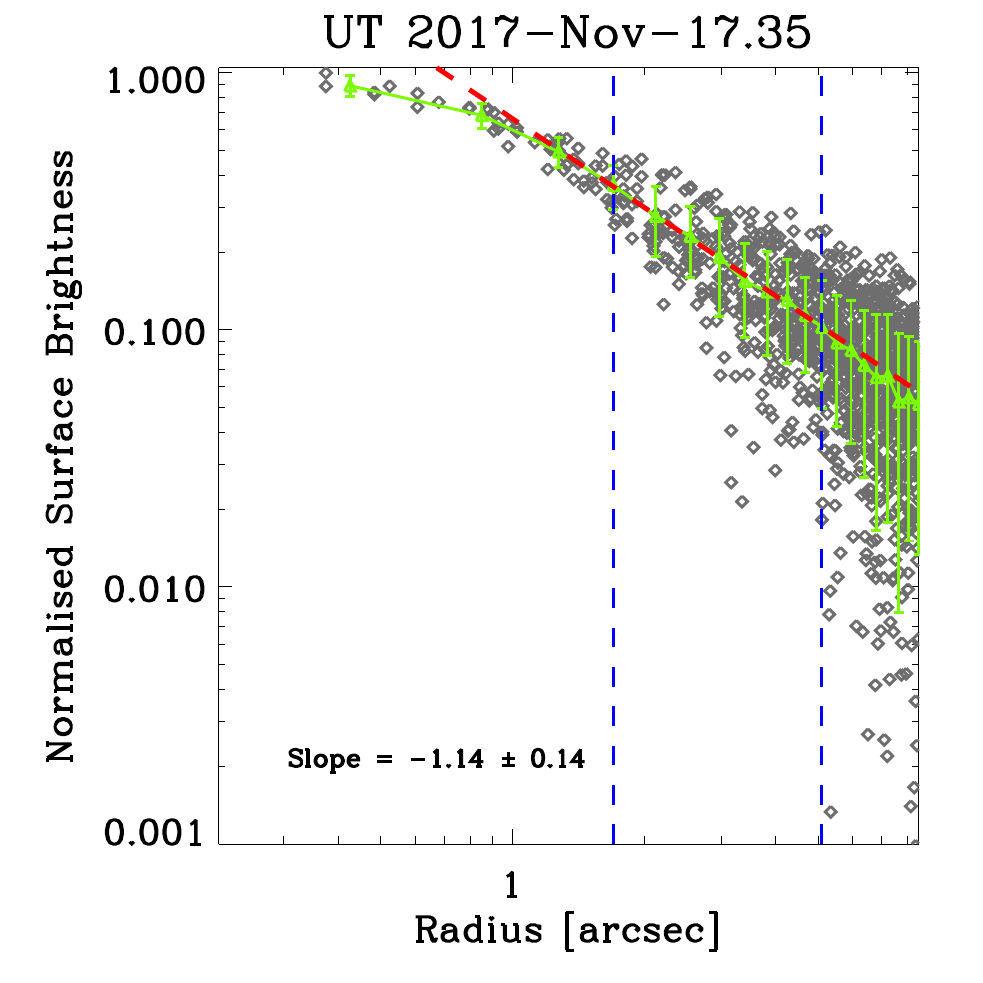}{0.32\textwidth}{(b)}
          \fig{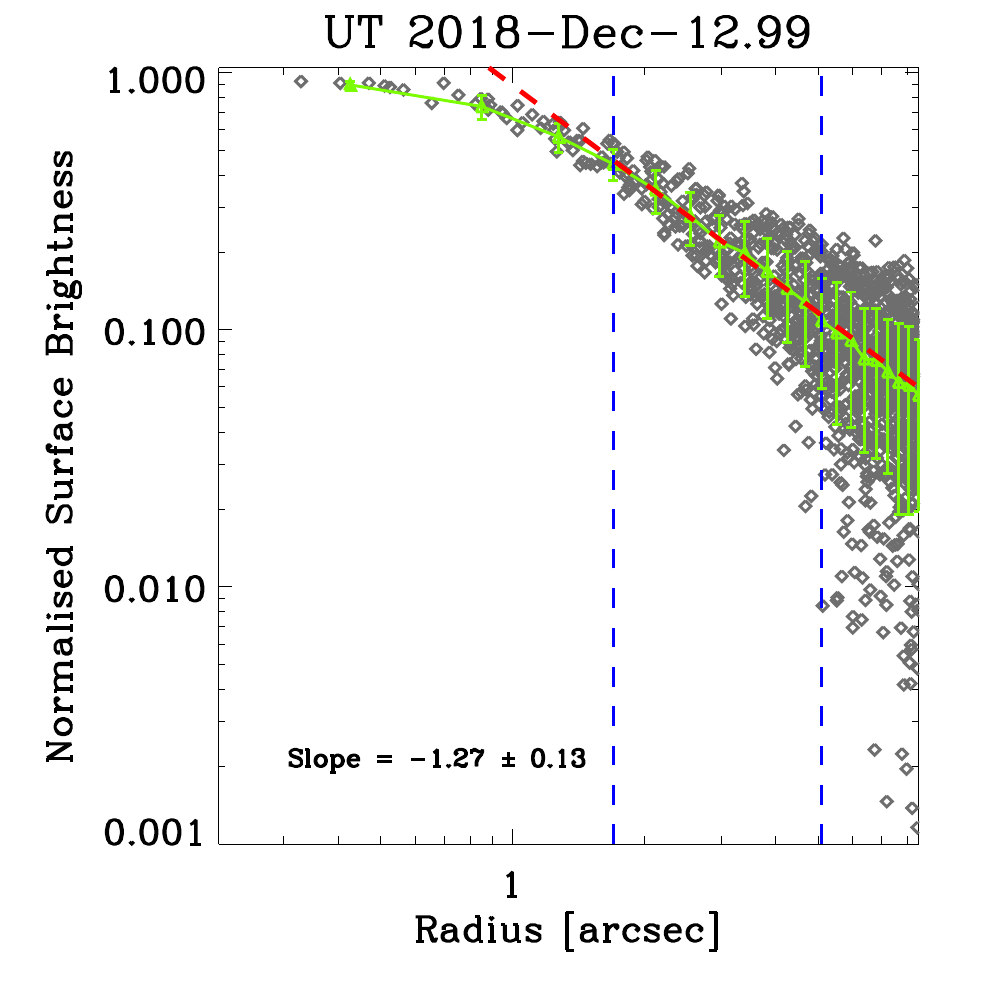}{0.32\textwidth}{(c)}
          }
\caption{
Examples of the radial brightness profiles of C/2010 U3 (Boattini) on 2011 January 30, 2017 November 17 and 2018 December 12, all normalized at the peak. In each panel, the azimuthal mean values of the counts are represented by the red dashed line, with the best fit shown in green, and the fit region is bounded by two vertical blue dashed lines.
\label{fig_U3_rsb}
} 
\end{figure*}

\section{Discussion}
\label{sec_disc}

\subsection{Morphology}
\label{subsec_morph}

We first compute the surface brightness profiles of U3 from three of the observation nights -- 2011 January 30, 2017 November 17 and 2018 December 12, when the neighbouring star fields were not crowded and seeing was superior (Figure \ref{fig_U3_rsb}). For each of the observations, the radii in pixels are first rounded to integers, and then counts in the same bins of radii are averaged with weights determined by the count uncertainties. The errors of the means are weighted standard deviations. We then fit the radial profiles in radius range 4-12 pixels from the optocentre by a power law. What will be characteristic is the logarithmic gradient ${\it \Gamma}$ of the coma. For a steady-state coma, ${\it \Gamma} = -1$, while ${\it \Gamma} = -1.5$ for a steady-state coma under the solar radiation pressure force \citep{1987AJ.....93.1542J}. Our obtained values of the logarithmic gradient are consistently $-1.5 < {\it \Gamma} < -1$, seemingly suggesting that the coma of U3 was in an intermediate state. However, the error bars are quite large as well. Therefore we opt not to further interpret it.

Nevertheless, the coma of comet U3 is apparently asymmetric, and a short tail has been seen since the earliest CFHT images on 2005 November 05 (see Figure \ref{fig_U3}). This is distinguished from comet K2, whose coma is circularly symmetric at similar heliocentric distances while the cometary activity in terms of effective scattering cross-section (see Section \ref{subsec_mloss} for U3) is comparable \citep{2017ApJ...847L..19J, 2018AJ....155...25H}.

\begin{figure*}
\epsscale{1.0}
\begin{center}
\plotone{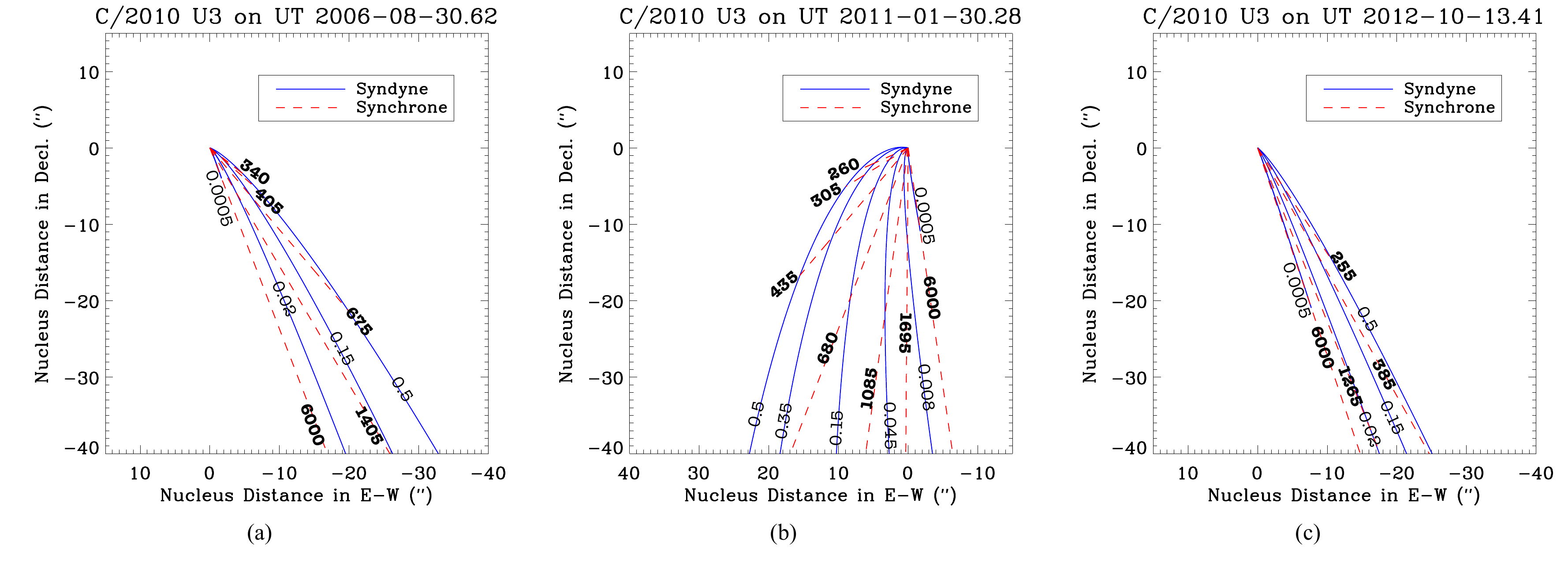}
\caption{
Examples of syndyne-synchrone grids for C/2010 U3 (Boattini). Only the solar gravitation and radiation pressure forces are considered. As indicated, syndynes are plotted as blue solid lines, and synchrones are red dashed lines. Also labelled are the values of the syndynes (adimensional, unbolded) and synchrones (in days, bolded). Obviously the observed orientation of the tail (Figure \ref{fig_U3}) is unmatched with that in these models.
\label{fig_U3_ss}
} 
\end{center} 
\end{figure*}

\begin{figure*}
\epsscale{1.0}
\begin{center}
\plotone{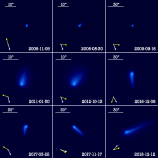}
\caption{
Modelled morphology of comet C/2010 U3 (Boattini) with inclusion of the Lorentz force. Dates in UT and scale bars are labelled in each panel. The yellow arrows mark the position angles of the antisolar directions, while the white ones mark the position angles of negative heliocentric velocity. Equatorial north is up and east is left. See Table \ref{tab_mdlpar} for the adopted parameters and Figure \ref{fig_U3} for the observations.
\label{fig_U3_mdl}
} 
\end{center} 
\end{figure*}

We analyse the tail first by applying the syndyne-synchrone computation based on \citet{1968ApJ...154..327F}. The model assumes that the dust grains leave the nucleus surface in zero initial velocity with respect to the nucleus, and then are subjected to the solar radiation pressure force and the local gravitational force due to the Sun, whose ratio, denoted as $\beta$, is related to dust grain properties by $\beta = \mathcal{C} Q_{\rm pr} \left(\rho_{\rm d} \mathfrak{a}\right)^{-1}$. Here $\rho_{\rm d}$ and $\mathfrak{a}$ are the bulk density and the radius of the dust grains, respectively, $\mathcal{C} = 5.95 \times 10^{-4}$ kg m$^{-2}$ is a proportionality constant, and $Q_{\rm pr}$ is the adimensional scattering efficiency assumed to be unity for the observed dust \citep[e.g.,][]{1979Icar...40....1B}. Positions of the grains are uniquely determined by the release time from the observed epoch ($\Delta \tau$) and $\beta$. A synchrone line is the loci of dust grains released at the same time from the nucleus with varying $\beta$, and a syndyne line is the loci of dust grains with same $\beta$ but released at various epochs.

Syndyne-synchrone grids for different epochs of the observations were computed. Our general conclusion is that the observed morphology cannot be explained by the syndyne-synchrone computation whatsoever, as the position angle of the tail cannot be formed by any combinations of the syndyne and synchrone lines (Figure \ref{fig_U3_ss}). Particularly, the CFHT observation from 2005 November 05 happened to be conducted at the exact epoch when the Earth was crossing the orbital plane of the comet (Table \ref{tab_vgeo}). As a result, all of the syndyne and synchrone lines would collapse into a single line when viewed from the Earth. However, the observation exhibits a faint tail pointing broadly northwards, unambiguously implying out-of-plane motion of the dust grains. 

We consider two possibilities that can give rise to the noted discrepancy: 1) the observed grains of U3 were ejected from the nucleus surface with nontrivial initial velocities in some preferential direction that has an out-of-plane component, and 2) the Lorentz force may have played an important role in diverting the trajectories of the dust particles.

Let us first investigate whether the morphology can be explained by nonzero initial velocities of the grains. The specific energy of a dust grain can be written as
\begin{align}
\nonumber
\mathscr{E}_{\rm d} & = \frac{1}{2} {\bf v}_{\rm d}^{2} - \frac{\left(1 - \beta \right) GM_{\odot}}{r_{\rm H}} \\
 & = \mathscr{E}_{\rm n} + \frac{1}{2}{\bf v}_{\rm ej}^{2} + {\bf v}_{\rm ej} \cdot {\bf v}_{\rm n} + \frac{\beta GM_{\odot}}{r_{\rm H}}
\label{eq_Ed}.
\end{align}
\noindent Here, $G = 6.67 \times 10^{-11}$ m$^{3}$ kg$^{-1}$ s$^{-2}$ is the gravitational constant, $M_{\odot} \approx 2 \times 10^{30}$ kg is the mass of the Sun, $\mathscr{E}$ is the specific energy with subscripts ``d" and ``n" representing for the dust and the cometary nucleus, respectively, $\mathbf{v}_{\rm n}$ is the heliocentric velocity of the nucleus, and $\mathbf{v}_{\rm ej}$ is the ejection velocity of the dust. To invalidate the synchrone-syndyne approximation, it is required that the third term in Equation (\ref{eq_Ed}) is at least comparable to the last term in the right-hand side, namely,
\begin{align}
\nonumber
\left|{\bf v}_{\rm ej} \right| & \gtrsim \frac{\beta G M_{\odot}}{r_{\rm H} \left|{\bf v}_{\rm n} \right|} \\
 & \gtrsim \beta \sqrt{\frac{G M_{\odot}}{2r_{\rm H}}}
\label{ineq_Vej},
\end{align}
\noindent in which we have applied the vis via equation to substitute $\left|\mathbf{v}_{\rm n} \right|$ with $r_{\rm H}$. Therefore, it will be unsurprising to see the syndyne-synchrone approximation fail at great heliocentric distances, as the ejection speed of the small-sized dust grains that are well coupled with the gas drag has a much weaker dependence upon $r_{\rm H}$.

We adopt our Monte Carlo dust ejection code based on the ejection model by \citet{2008Icar..193...96I}, in which the initial ejection velocities of dust grains and perturbation from the eight major planets are considered in addition to the solar gravitational force and the radiation pressure force, to simulate the observed morphology of U3. Similar analyses have been previously applied to other long-period comets as well \citep[e.g.,][]{2014ApJ...787..115Y}. After examining a series of parameters including the initial ejection velocities and the sizes of the dust grains, we realize that the observed morphology cannot be accounted for any dust ejection near the subsolar point of the nucleus, but nearly 90\degr~away from it. If the activity of U3 is continuous, ejection with such an obvious deviation from the subsolar point will be unphysical, because at such great heliocentric distances, we expect that temperature gradients across the nucleus will be milder, resulting in only a small angle between the local maximum temperature spot and the subsolar point, if at all. 

We also reject the possibility that the morphology could be formed by a single or multiple fragmentation events. Assuming the fragmentation occurs exactly at the earliest CFHT observation of U3 on 2005 November 05, given a typical separation speed of $\sim$0.1-1 m s$^{-1}$ \citep{1982come.coll..251S}, we should be able to effortlessly notice the disconnection between the nucleus and the debris cloud of $\gtrsim$7\arcsec~in width in the latest WIYN observation of U3. However, we see no such evidence. We also attempt to fit the astrometry with a nongravitational force model following $r_{\rm H}^{-2}$, but fail to obtain a significant radial nongravitational parameter ($< 1\sigma$), implying that what we have been observing about U3 cannot be a debris cloud either.

\begin{deluxetable}{cccc}
\tablecaption{Adopted Parameter Values for Morphology Simulation of Comet C/2010 U3 (Boattini)
\label{tab_mdlpar}}
\tablewidth{0pt}
\tablehead{
\colhead{Date (UT)} & \colhead{$\mathfrak{a}_{\min}$ (\micron)\tablenotemark{\dag}} & \colhead{$\Delta \tau_{\max}$ ($10^{7}$ s)\tablenotemark{\ddag}} & 
\colhead{$\left|\bf{v}_{\rm ej, 0} \right|$ (m s$^{-1}$)\tablenotemark{$\ast$}}
}
\startdata
2005 Nov 05 & 10 & 2.5 & 2.0 \\
2006 Aug 30 & 10 & 3.0 & 2.0 \\
2009 Sep 16 & 10 & 2.5 & 2.0 \\
2011 Jan 30 & 5 & 2.0 & 5.0 \\
2012 Oct 13 & 5 & 2.0 & 5.0 \\
2016 Dec 09 & 10 & 2.0 & 4.0 \\
2017 Mar 25 & 20 & 3.0 & 3.0 \\
2017 Nov 17 & 4 & 1.0 & 3.0 \\
2018 Dec 12 & 5 & 1.5 & 5.0 \\
\enddata
\tablenotetext{\dag}{Minimum dust grain radius.}
\tablenotetext{\ddag}{Maximum dust ejection time, measured from the observed epoch, positive going backward.}
\tablenotetext{\ast}{Referenced to grains of 5 mm in radius ejected at $r_{\rm H} = 1$ au.}
\tablecomments{
We have adopted common $\mathfrak{a}_{\max} = 1$ mm and $\Delta \tau_{\min} = 0$ s for all of the epochs. See Figure \ref{fig_U3_mdl} for the simulated morphology of the comet. Following \citet{2008Icar..193...96I}, the ejection terminal speed is assumed to be correlated with the dust size and heliocentric distance as $\left| {\bf v}_{\rm ej} \right| \propto \left(\mathfrak{a} r_{\rm H} \right)^{-1/2}$. We also restricted the ejection of dust to be within $\sim$10\degr~from the subsolar point, forming a cone-shape jet. If the active source is wider, e.g., all over the sunlit hemisphere, the tail will be much wider than what the observations showed. Conversely, if the cone is too narrow, e.g., $\sim$1\degr, the tail will seem too narrow. Nevertheless, our adopted parameters appear to reproduce the observed morphology of the comet the best, and they are order-of-magnitude comparable at different epochs.
}
\end{deluxetable}

Therefore, we start to explore whether the Lorentz force can be at play. At similar heliocentric distance, comet C/1995 O1 (Hale-Bopp) exhibited a tail which was probably diverted by the Lorentz force observably \citep{2014Icar..236..136K}. In interplanetary space, a dust grain is charged to a positive surface potential of $\mathscr{U} \approx +5$ V due to the loss of photoelectrons by solar UV \citep[e.g.,][]{1998ApJ...499..454K}. Assuming a spherical shape, such a grain will carry a charge of $\mathscr{Q} = 4\pi \epsilon_{0} \mathscr{U} \mathfrak{a}$, where $\epsilon_0 = 8.85 \times 10^{-12}$ F m$^{-1}$ is the permittivity of free space. Hence, the charge-to-mass ratio of the grain is related to $\beta$ by
\begin{equation}
\frac{\mathscr{Q}}{\mathfrak{m}_{\rm d}} = \frac{3 \epsilon_{0} \mathscr{U} \rho_{\rm d}}{\mathcal{C}^2 Q_{\rm pr}^2} \beta^2
\label{eq_q2m}.
\end{equation}

 Similar to previous works \citep[e.g.,][]{2012ApJ...744..170P, 2013GeoRL..40.2500J}, we approximate the interplanetary magnetic field strength ${\bf B}$ by Parker's spiral \citep{1958ApJ...128..664P}:
\begin{equation}
{\bf B} = \pm B_{\rm R,0} \left(\frac{r_\oplus}{r_{\rm H}} \right)^2
\left[1 - 2 \mathscr{H} \left(\mathscr{B}_{\rm cs} - \mathscr{B} \right) \right]
\begin{pmatrix}
1  \\
-\dfrac{\Omega_\odot r_{\rm H}}{\left| {\bf v}_{\rm sw} \right|} \cos \mathscr{B} \\
0
\end{pmatrix}
\label{eq_B}.
\end{equation}
\noindent Here, the expression is referenced to the heliographic RTN system, $B_{\rm R,0} = 3$ nT is the radial component of the magnetic field strength at the mean Sun-Earth distance $r_\oplus = 1$ au, $\mathscr{B}$ and $\mathscr{B}_{\rm cs}$ are the heliographic latitudes of the dust and the current sheet (CS), respectively, $\mathscr{H}$ is the Heaviside function, and $\Omega_\odot = 0.248$ day$^{-1}$ is the sidereal spin rate of the Sun. The polarity reversal is approximated to occur and complete instantaneously whenever the solar maximum is reached halfway in some solar cycle. When the radial magnetic fields emanate from the southern hemisphere of the Sun, the minus sign in Equation (\ref{eq_B}) is taken. We compute time-dependent $\mathscr{B}_{\rm cs}$ from the potential field source surface model (\url{http://wso.stanford.edu/Tilts.html}). For the solar wind, we assume that its speed is $\left| {\bf v}_{\rm sw} \right| = 750$ km s$^{-1}$ for $\left|\mathscr{B}\right| \gtrsim 20\degr$, and $\left| {\bf v}_{\rm sw} \right| = 400$ km s$^{-1}$ otherwise, which is globally in line with measurements from the Ulysses spacecraft \citep{1995GeoRL..22.3301P}. With Equations (\ref{eq_q2m}) and (\ref{eq_B}), the acceleration of the dust grain due to the Lorentz force, 
\begin{equation}
{\bf a}_{\rm L} = \frac{\mathscr{Q}}{\mathfrak{m}_{\rm d}} \left({\bf v}_{\rm d} - {\bf v}_{\rm sw} \right) \times {\bf B}
\label{eq_aL},
\end{equation}
\noindent can be expressed, which is then transformed to the ecliptic coordinate system and added as an additional perturbation source in our Monte Carlo dust ejection code.

We find that the model with incorporation of the Lorentz force and broadly similar sets of parameters (Table \ref{tab_mdlpar}) can successfully reproduce the morphology of U3 at all of the observed epochs (Figure \ref{fig_U3_mdl}). The fluctuation in the minimum dust size $\mathfrak{a}_{\min}$ probably hints at intrinsic variations in the activity of the comet, or variations in the interplanetary magnetic field. We cannot confine the maximum dust-grain size so simply assume $\mathfrak{a}_{\max} = 1$ mm. Another parameter we assume is the power-law index of the differential dust-size distribution, $\gamma = -3.5$, after checking that $\gamma$ does not affect the morphology or the surface brightness profile strongly \citep[also see Figure 6 in][]{2008Icar..193...96I}. Nevertheless, we deduce that the observed dust grains have minimum radii of $\mathfrak{a}_{\min} \sim 10$ \micron, and were ejected from the nucleus protractedly at speeds of $\left| {\bf v}_{\rm ej} \right| \lesssim$ 50 m s$^{-1}$ within a cone symmetric about the Sun-comet axis at the subsolar point whose half-opening angle is $\sim$10\degr. 

Hitherto, there has been only sparse evidence of cometary dust under strong influence by the Lorentz force due to the fact that we are strongly biased towards comets that are much closer to the Sun ($r_{\rm H} \lesssim 5$ au), where the solar radiation pressure force is typically dominant. To see this, we express the ratio between the solar radiation pressure force and the Lorentz force of the dust as
\begin{equation}
\beta_{\rm L} = \frac{\mathcal{C} Q_{\rm pr} G M_\odot}{3\epsilon_0 \mathscr{U} B_{\rm R,0} r_{\oplus}^2 \Omega_\odot \cos \mathscr{B}} \left( \frac{\mathfrak{a}}{r_{\rm H}} \right)
\label{eq_beta_L}.
\end{equation}
\noindent The Lorentz force begins to rival the radiation counterpart if $\beta_{\rm L} < 1$ is satisfied. At low heliographic latitudes, this can be easily satisfied by particles of, for instance, $\mathfrak{a} \lesssim 0.5$ \micron~at $r_{\rm H} = 10$ au, and $\mathfrak{a} \lesssim 1$ \micron~at $r_{\rm H} = 20$ au, both efficient in scattering sunlight in the optical wavelengths. We foresee that more samples of comet morphologies influenced by the Lorentz force for ultra-distant comets will be identified in the near future, thanks to our improving capacity in detecting distant comets.

\begin{deluxetable*}{lcccc}
\tablecaption{Normalized Reflectivity Gradients of Comet C/2010 U3 (Boattini)
\label{tab_sprime}}
\tablewidth{0pt}
\tablehead{
\colhead{Date} & \colhead{Telescope} & \colhead{$S'\left(B,V\right)$} & \colhead{$S'\left(V,R \right)$} & \colhead{$S'\left(B,R\right)$} \\
\colhead{(UT)} & & \colhead{(\% per $10^3$ \AA)} & \colhead{(\% per $10^3$ \AA)} & \colhead{(\% per $10^3$ \AA)}
}
\startdata
2006 Aug 24\tablenotemark{\dag} & CFHT & $12.6 \pm 11.2$ & $11.6 \pm 10.2$ & $12.1 \pm 7.6$\\
2009 Sep 16 & SDSS & $10.0 \pm 10.5$ & $9.6 \pm 10.4$ & $9.8 \pm 7.4$\\
2011 Jan 30 & Keck & $10.2 \pm 3.0$ & $7.9 \pm 2.1$ & $9.1 \pm 1.6$ \\
2012 Oct 13 & Keck & $13.1 \pm 5.7$ & $13.3 \pm 6.9$ & $13.1 \pm 3.5$\\
2012 Oct 14 & Keck & $11.4 \pm 5.4$ & $18.6 \pm 7.7$ & $14.6 \pm 3.4$\\
2016 Dec 09 & WIYN & $30.8 \pm 7.4$ & $6.5 \pm 3.4$ & $19.2 \pm 4.0$ \\
2017 Mar 25 & WIYN & $27.1 \pm 2.8$ & $8.2 \pm 4.5$ & $18.0 \pm 2.2$\\
2017 Nov 14 & WIYN & N/A & $11.3 \pm 5.1$ & N/A \\
2017 Nov 17 & WIYN & $23.4 \pm 4.3$ & $15.0 \pm 3.6$ & $19.2 \pm 2.7$\\
2018 Dec 09 & WIYN & $14.4 \pm 3.8$ & $13.9 \pm 3.4$ & $14.1 \pm 1.9$\\
2018 Dec 12 & WIYN & $20.0 \pm 4.7$ & $12.5 \pm 2.5$ & $16.3 \pm 2.3$\\
2018 Dec 13 & WIYN & $14.4 \pm 8.6$ & $14.6 \pm 4.9$ & $14.4 \pm 4.0$\\
\enddata
\tablenotetext{\dag}{We combined the CFHT observations from 2006 August 18 and 30 with assumption that the absolute magnitude of the comet remains unaltered to derive the {\it g}' $-$ {\it r}' color index, which is then transformed to color indices in the Johnson-Cousins system using conversion equations by \citet{Jordi06}. }
\tablecomments{
Uncertainties in $S'\left(\lambda_1, \lambda_2 \right)$ are propagated from the magnitude errors.
}
\end{deluxetable*}

\begin{figure}
\gridline{\fig{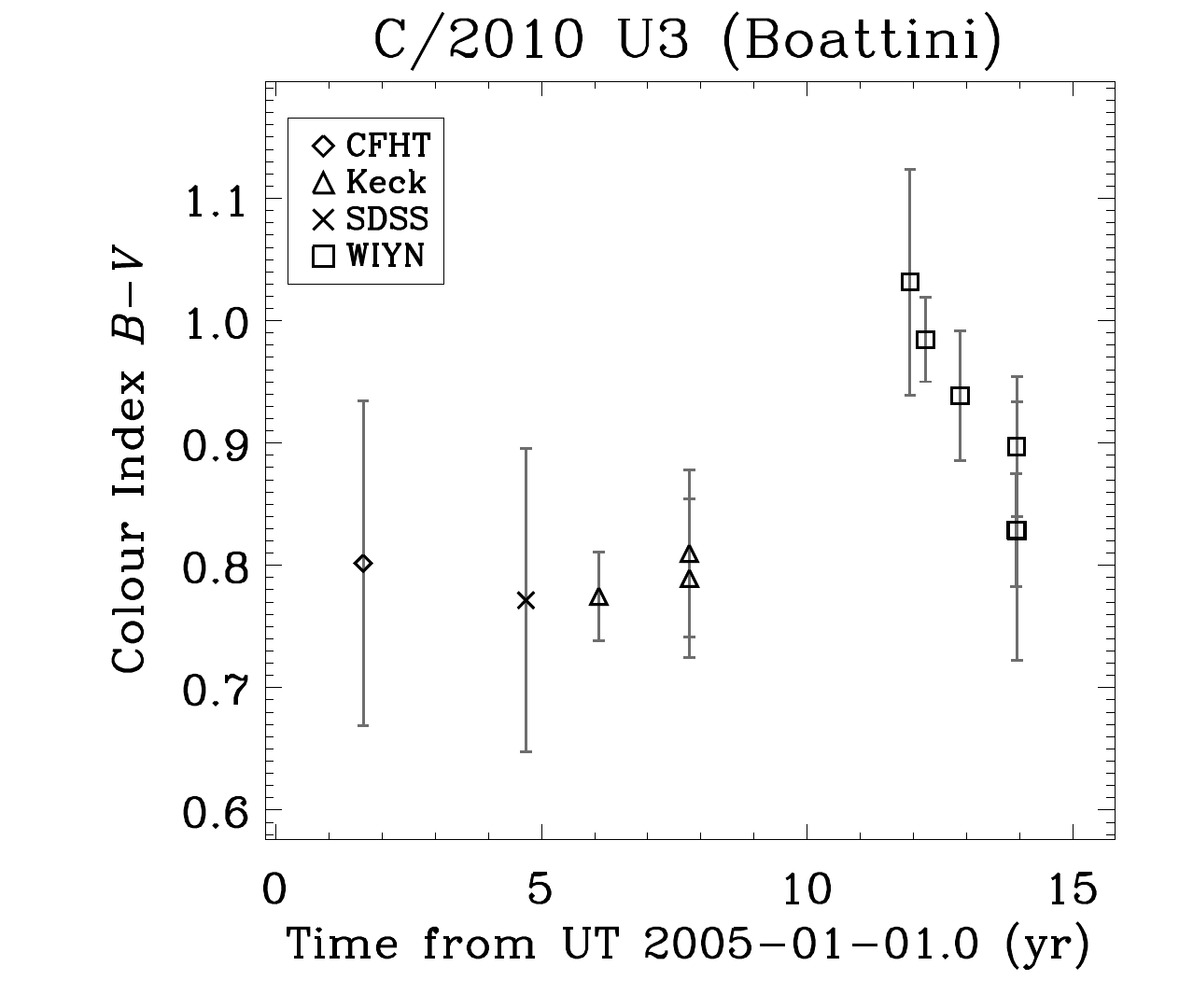}{0.4\textwidth}{(a)}
          }
\gridline{\fig{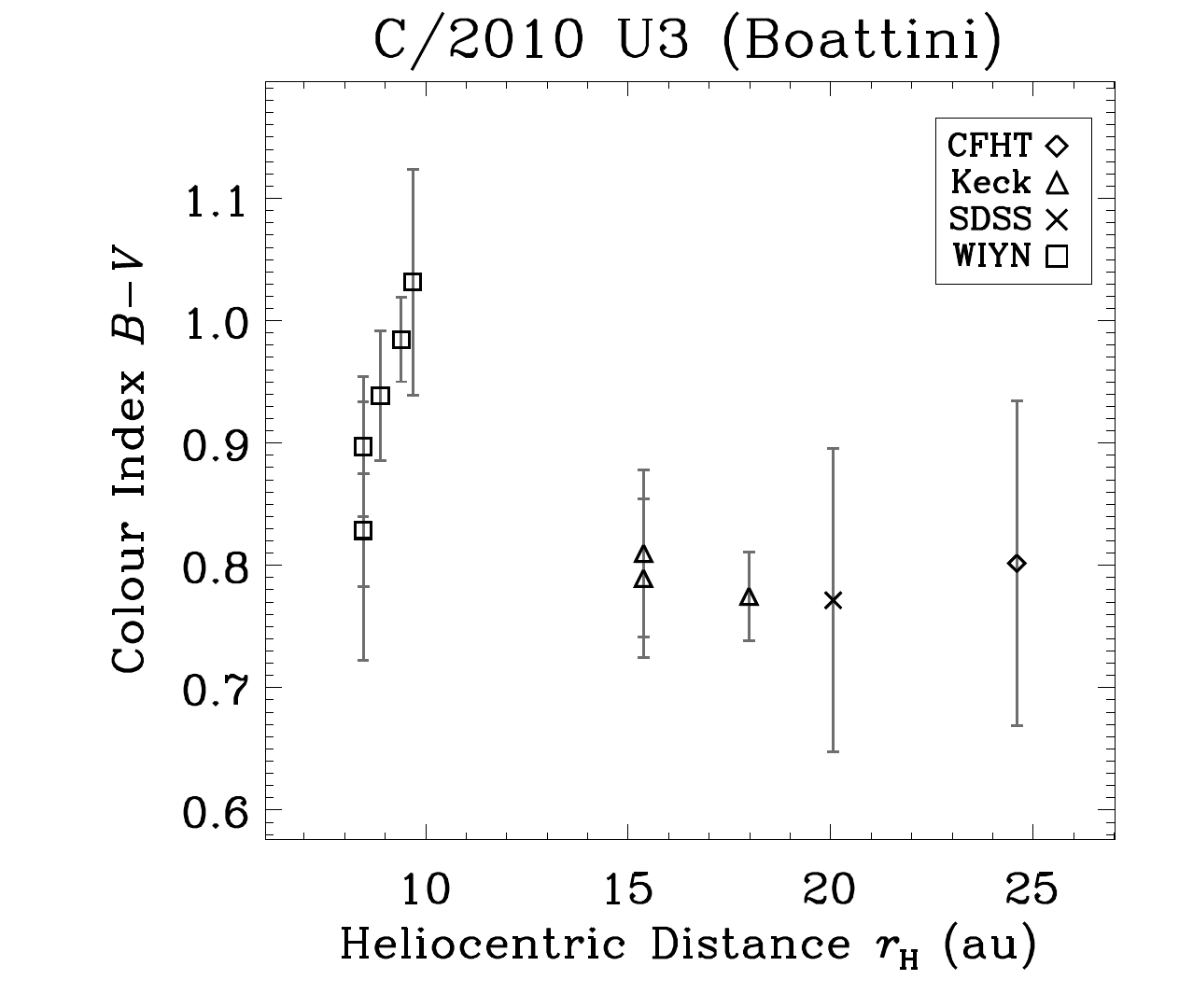}{0.4\textwidth}{(b)}
          }
\caption{
The {\it B} $-$ {\it V} color index of comet C/2010 U3 (Boattini) as functions of time (a) and heliocentric distance (b). Observations in the Sloan system have been transformed to the Johnson-Cousins system using conversion equations derived by \citet{Jordi06}. Data points from different observatories are discriminated by the symbols labelled in the legend. Note that the comet appeared the reddest in the {\it B} $-$ {\it V} wavelength interval when the earliest WIYN observations were made, but then gradually blued and restored the original color.
\label{fig_U3_BV}
} 
\end{figure}

\begin{figure}
\gridline{\fig{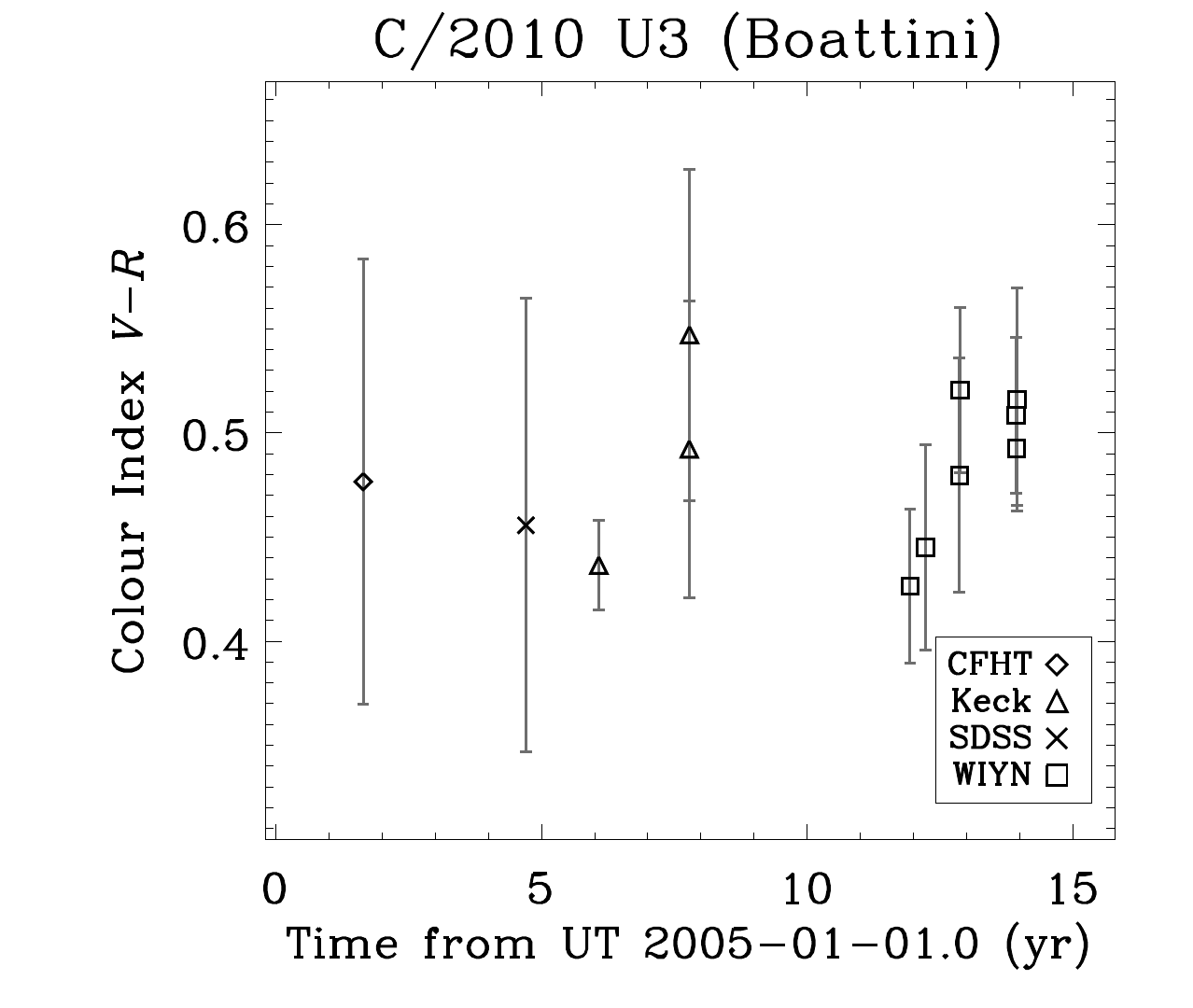}{0.4\textwidth}{(a)}
          }
\gridline{\fig{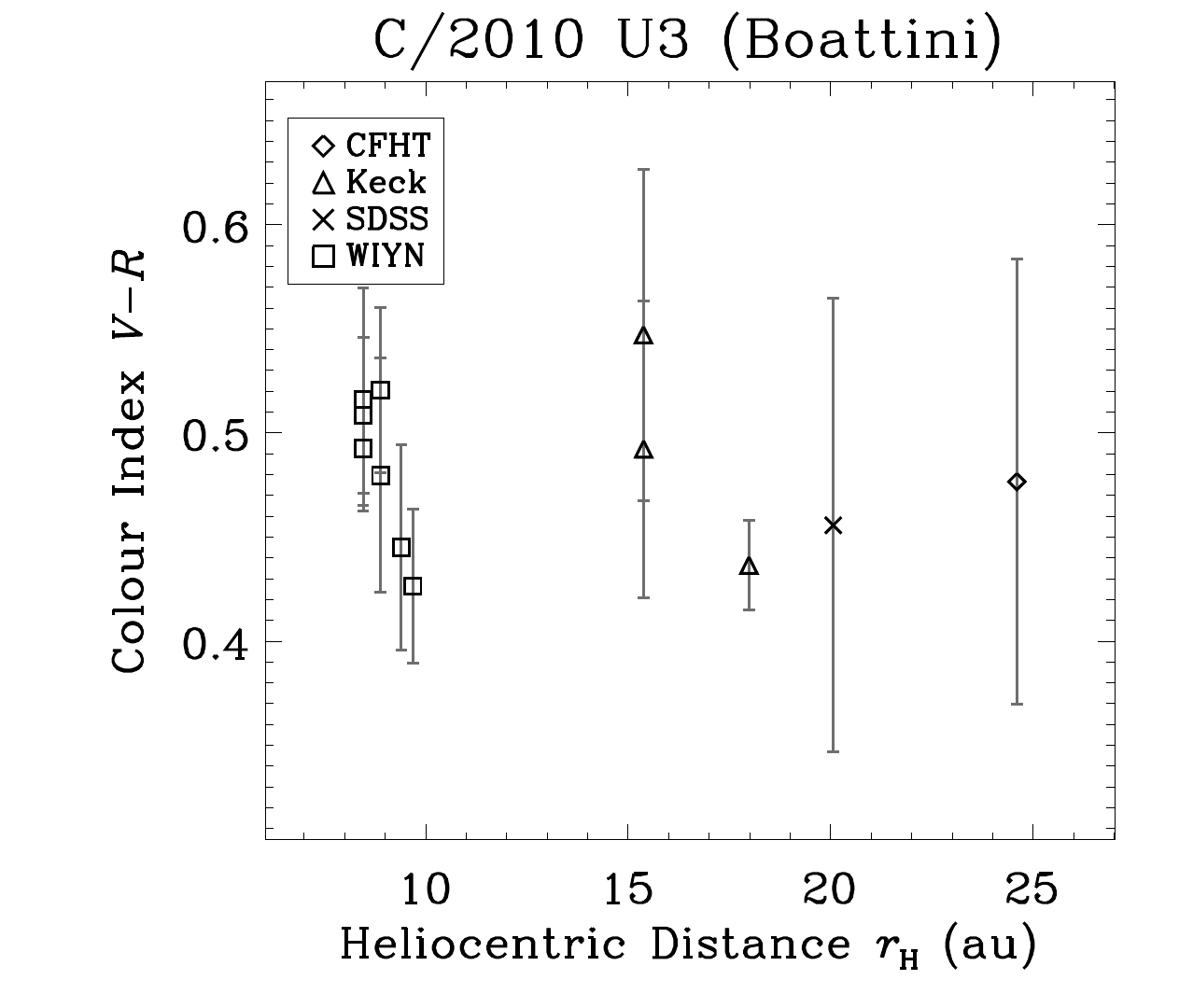}{0.4\textwidth}{(b)}
          }
\caption{
Same as Figure \ref{fig_U3_BV}, but for the {\it V} $-$ {\it R} color index as functions of time and heliocentric distance. It appears that, opposite to what we see in the {\it B} $-$ {\it V} wavelength interval, the comet potentially experienced reddening in the {\it V} $-$ {\it R} interval since our first WIYN observation of the comet.
\label{fig_U3_VR}
} 
\end{figure}

\begin{figure}
\epsscale{1.0}
\begin{center}
\plotone{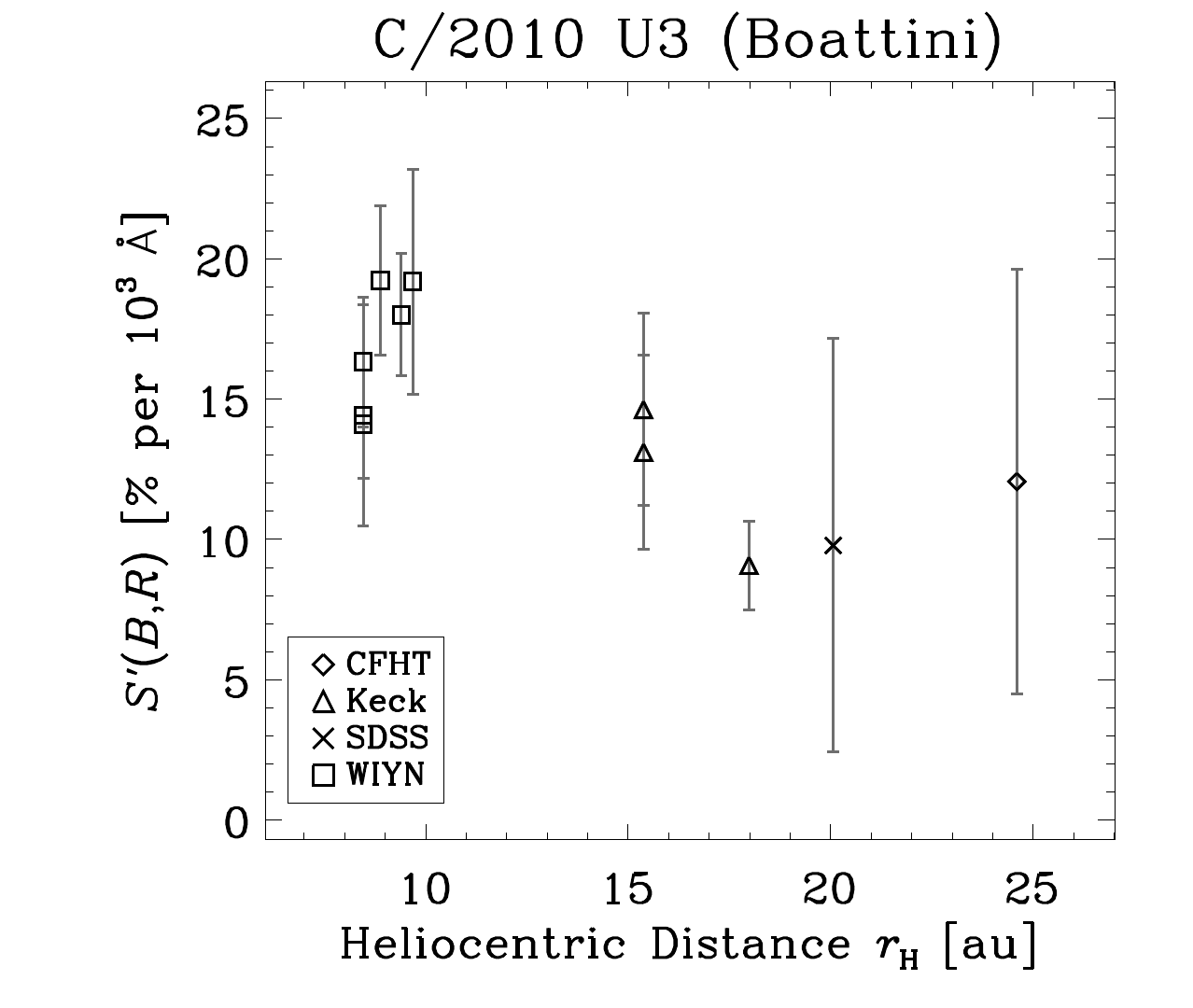}
\caption{
The normalized reflectivity gradient of comet C/2010 U3 (Boattini) in the {\it B} $-$ {\it R} wavelength interval as a function of heliocentric distance. Data points from different observatories are distinguished by symbols. Associated errors are propagated from errors in the photometry measurements with Equation (\ref{eq_sprime}). 
\label{fig_U3_S_BR}
} 
\end{center} 
\end{figure}

\subsection{Color}
\label{subsec_clr}

We plot the color indices {\it B} $-$ {\it V} and {\it V} $-$ {\it R} of U3 as functions of time and heliocentric distance in Figures \ref{fig_U3_BV} and \ref{fig_U3_VR}, respectively. Noteworthily, the comet appeared redder in the {\it B} $-$ {\it V} interval at some point between the 2012 Keck and 2016 WIYN observations, at $10 < r_{\rm H} < 15$ au from the Sun, but then gradually turned bluer and restored the original color. In the {\it V} $-$ {\it R} interval, a possible opposite trend is seen. We therefore suspect that the comet intrinsically brightened in the {\it V} band momentarily. Intriguingly, the occurrence of the color change appeared to coincide with the onset of crystallisation of amorphous water ice on U3, if at all (see Section \ref{subsec_mloss}). However, we cannot be sure whether or not it was crystallisation of amorphous ice that caused the observed color variation.

For completeness, we also compute the normalized reflectivity gradient $S' \left(\lambda_1, \lambda_2 \right)$ defined by \citet{1984AJ.....89..579A} and \citet{1986ApJ...310..937J} as
\begin{equation}
S'\left( \lambda_1, \lambda_2 \right) = \left( \frac{2}{\lambda_{2} - \lambda_{1}} \right) \frac{10^{0.4 \left[ \Delta m_{1,2} - \Delta m_{1,2}^{(\odot)}\right] } - 1}{10^{0.4 \left[ \Delta m_{1,2} - \Delta m_{1,2}^{(\odot)}\right] } + 1}
\label{eq_sprime},
\end{equation}
\noindent where $\Delta m_{1,2}$ and $\Delta m_{1,2}^{(\odot)}$ are respectively the color indices of the comet and the Sun in the filter pair. Conveniently, if an object has essentially the same color as that of the Sun, Equation (\ref{eq_sprime}) yields $S'\left(\lambda_1, \lambda_2 \right) = 0$. Colors redder than that of the Sun correspond to $S'\left(\lambda_1, \lambda_2 \right) > 0$, otherwise $S'\left(\lambda_1, \lambda_2 \right) < 0$. The results are presented in Table \ref{tab_sprime} and plotted in Figure \ref{fig_U3_S_BR}, where we can see that the color of U3 in the {\it B} $-$ {\it R} wavelength interval likely reddened during the observed period. Similar behaviors have been observed for another long-period comet C/2013 A1 (Siding Spring), which most likely suggests the existence of grains mixed with volatile ice and refractory material \citep{2014ApJ...797L...8L}.

We are aware that a possible bias in the measured color may be caused by inconsistent star catalogues used for photometric calibration (see Section \ref{sec_obs}). In order to rule out this possibility, we applied completely the same procedures to derive zero-points of several other WIYN images covered by both the SDSS DR12 and the PS DR1 in the Johnson-Cousins bands. Each image has a number of comparison stars similar to those of the WIYN images (at least a few tens). What we found is that the differences in zero-points between the two catalogue sources never exceed $\sim$$1.5\sigma$ of the zero-point errors ($\lesssim$0.05 mag). Therefore, we do not think that the color variation of U3 can be explained by our choice of different catalogue sources for photometry, but is authentic. 

Optically dominant dust grains at speeds of $\sim$10 m s$^{-1}$ need to spend $\sim$40 days crossing the photometric aperture, which is obviously shorter than the gaps between the observations of different runs. Thus, we observed generally different dust grains during each run, which indicates the chemical heterogeneity of the coma of U3, or different activity mechanisms in different heliocentric distance regimes, or perhaps both.

\begin{figure}
\epsscale{1.0}
\begin{center}
\plotone{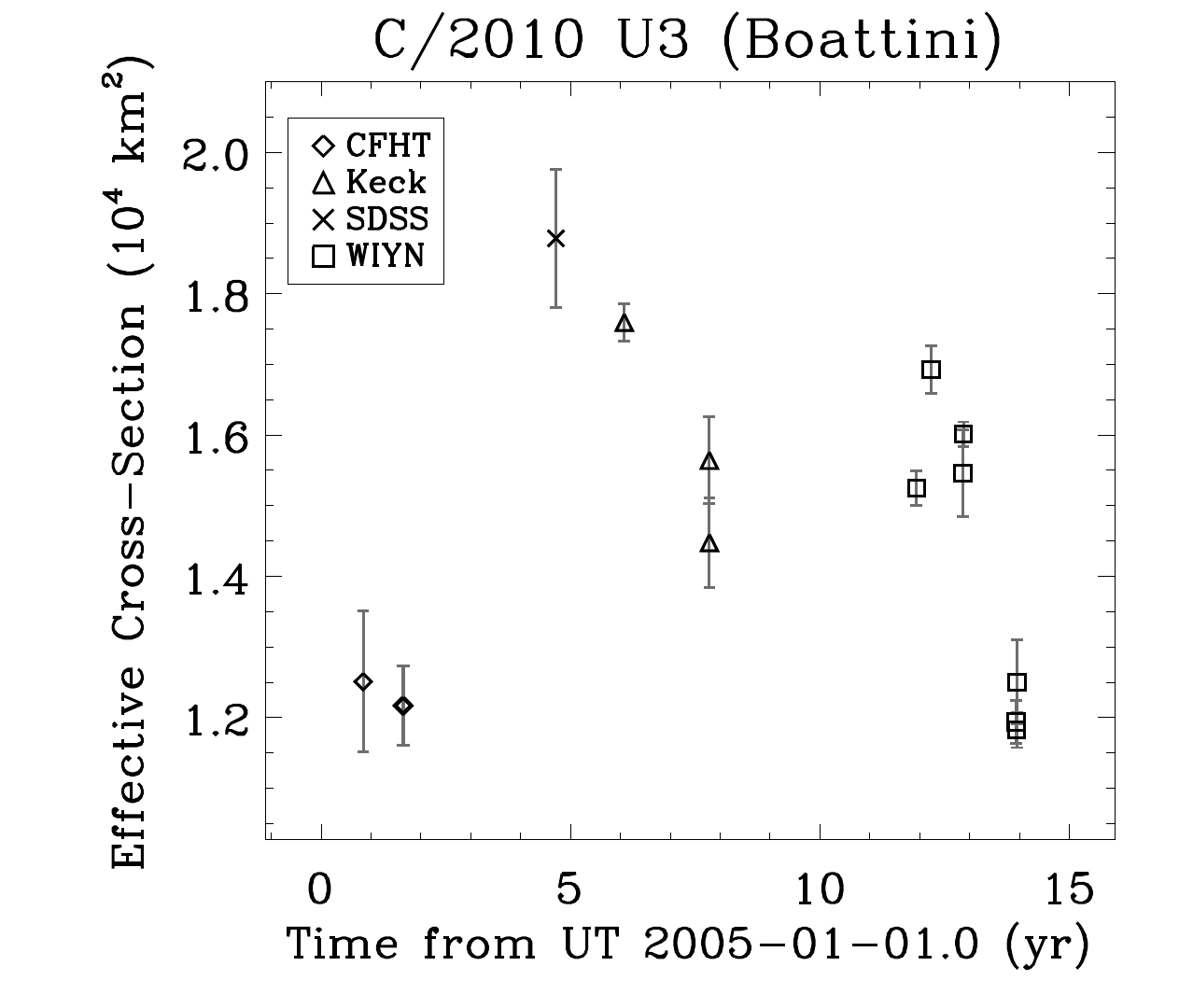}
\caption{
The effective scattering cross-section of comet C/2010 U3 (Boattini) as a function of time. Data point symbols correspond to observatories. Uncertainties are propagated from errors in the photometry measurements. The {\it V}-band geometric albedo is assumed to remain constant throughout the observed interval, $p_{V} = 0.04$, a typical value for cometary dust \citep{2004come.book..223L}.
\label{fig_U3_XS}
} 
\end{center} 
\end{figure}

\subsection{Activity Mechanism}
\label{subsec_mloss}

We estimate the effective scattering cross-section of U3, $C_{\rm e}$, from the absolute {\it V}-band magnitude using
\begin{equation}
C_{\rm e} = \frac{\pi r_{\oplus}^{2}}{p_{V}} 10^{0.4 \left[m_{\odot,V} - m_{V} \left(1,1,0 \right) \right]}
\label{eq_XS}.
\end{equation}
\noindent Here $m_{\odot, V} = -26.74$ is the apparent {\it V}-band magnitude of the Sun, and $p_{V}$ is the {\it V}-band geometric albedo of the comet, which is unknown. We thus assume a constant $p_{V} = 0.04$, typical for surfaces of cometary nuclei \citep{2004come.book..223L}. The resulting values of $C_{\rm e}$ are summarized in Table \ref{tab_phot} and plotted in Figure \ref{fig_U3_XS}.

The fluctuations in the effective scattering cross-section of U3 (Figure \ref{fig_U3_XS}) are most likely an indication of changes in the activity of the comet. When the inflow of dust grains in the photometric aperture outnumbers the outflow, a surge in the cross-section will be witnessed. Otherwise it will be a decline. Provided an optically thin coma, the net mass-loss rate is related to the effective cross-section by
\begin{equation}
\dot{\mathcal{M}} = \frac{4}{3} \rho_{\rm d} \bar{\mathfrak{a}} \dot{C}_{\rm e}
\label{eq_mloss}.
\end{equation}
\noindent Here, $\bar{\mathfrak{a}} = \sqrt{\mathfrak{a}_{\min} \mathfrak{a}_{\max}}$ is the mean grain radius. However, since we cannot constrain $\mathfrak{a}_{\max}$ effectively, we prefer a lower limit to the magnitude by substituting $\bar{\mathfrak{a}}$ with $\mathfrak{a}_{\min} \sim 10$ \micron. Inserting the values, we obtain $\overline{\dot{\mathcal{M}}} \gtrsim 0.5$ kg s$^{-1}$ for U3 during the period of 2006-2009 when it brightened, and $\overline{\dot{\mathcal{M}}} \lesssim -0.7$ kg s$^{-1}$ since early 2017 as the intrinsic brightness declined. Both estimates are probably accurate to order of magnitude at best. These crude lower limits are nevertheless noticeably smaller than that of comet K2 \citep[$\sim$200 kg s$^{-1}$;][]{2018AJ....155...25H} by two orders of magnitude. The main difference is that, although the two comets had comparable effective scattering cross-sections, the dust grains ejected from K2 are larger by approximately two orders of magnitude. 

The prolonged activity of comet U3 implies that sublimation of volatiles is the leading explanation. Temperatures at U3 are too low at the observed heliocentric distances such that only substances more volatile than water ice, e.g., CO and CO$_{2}$, would be able to sublimate. As the insolation power from the Sun is received at the nucleus, it will then be turned into powers for reradiation in the infrared and sublimation of volatiles. This can be expressed by the following equation:
\begin{equation}
\left(1 - A_{\rm B} \right) S_{\odot} \left(\frac{r_{\oplus}}{r_{\rm H}} \right)^2 \cos \zeta = \epsilon \sigma T^4 + L\left(T\right) f_{\rm s} \left(T\right)
\label{eq_E},
\end{equation}
\noindent where $A_{\rm B}$ is the Bond albedo, $S_{\odot} = 1361$ W m$^{-2}$ is the solar constant, $\cos \zeta$ is the effective projection coefficient for the surface ($1/4 \le \cos \zeta \le 1$, the lower and upper limits correspond to the isothermal and subsolar cases, respectively), $\epsilon$ is the emissivity, $\sigma = 5.67 \times 10^{-8}$ W m$^{-2}$ K$^{-4}$ is the Stefan-Boltzmann constant, $T$ is the surface temperature, $L$ is the latent heat of the sublimating substance, and $f_{\rm s}$ is its mass flux. The heat conduction towards the nucleus interior from the surface is ignored, in that the thermal conductivity of cometary nuclei is believed to be tiny \citep[e.g.,][]{2006hgdc.conf.....H}. To solve Equation (\ref{eq_E}), we assign $A_{\rm B} = 0.01$ and $\epsilon = 0.9$, which are typical values for cometary nuclei \citep[e.g.,][]{2004Icar..167...16B}, since we have no pertinent knowledge. We adopt empirical thermodynamic parameters of CO and CO$_2$ as representatives of supervolatiles respectively from \citet{2006hgdc.conf.....H} and \citet{1979M&P....21..155C}.

Equation (\ref{eq_E}) is solved numerically. We obtain that, for sublimation of CO, the mass flux rises from $1.3 \times 10^{-6} \lesssim f_{\rm s} \lesssim 5.6 \times 10^{-6}$ kg s$^{-1}$ m$^{-2}$ in late 2005, to $1.3 \times 10^{-5} \lesssim f_{\rm s} \lesssim 5.5 \times 10^{-5}$ kg s$^{-1}$ m$^{-2}$ in late 2018, where the lower and upper ends correspond to the isothermal and subsolar scenarios, respectively. In comparison, for CO$_2$, it increases from $4.9 \times 10^{-15} \lesssim f_{\rm s} \lesssim 2.5 \times 10^{-8}$ kg s$^{-1}$ m$^{-2}$ to $2.3 \times 10^{-6} \lesssim f_{\rm s} \lesssim 2.5 \times 10^{-5}$ kg s$^{-1}$ m$^{-2}$. To sustain the observed net mass-loss rate of the comet, a minimum active surface area is needed:
\begin{equation}
\mathcal{A}_{\rm s} = \frac{\left| \dot{\mathcal{M}} \right|}{\mathcal{X} f_{\rm s}}
\label{eq_As},
\end{equation}
\noindent in which $\mathcal{X}$ is the dust-to-gas mass ratio. The majority of comets have $\mathcal{X} < 2$ \citep{1992AJ....104..848S,1996A&AS..120..301S}, however, there apparently exist exceptions such as C/1995 O1 (Hale-Bopp) and C/2011 L4 (PANSTARRS), whose dust-to-gas mass ratios are as high as $\mathcal{X} \gtrsim 4$ \citep{1999AJ....117.1056J,2014ApJ...784L..23Y}. Assuming a typical value of $\mathcal{X} = 1$ for U3, we find that its minimum active surface area is $0.1 \lesssim \mathcal{A}_{\rm s} \lesssim 0.2$ km$^{2}$ for CO as the main sublimating substance, and $1 \lesssim \mathcal{A}_{\rm s} \lesssim 7 \times 10^{5}$ km$^2$ for the CO$_2$ case. These respectively correspond to equal-area circle of radii $0.1\lesssim R_{\rm n} \lesssim 0.3$ km and $0.6 \lesssim R_{\rm n} \lesssim 500$ km.

We can therefore see that while sublimation of CO (or substances of similar volatility, e.g., N$_2$, O$_2$) can easily suffice the net mass-loss rate of U3, sublimation of CO$_2$ will be needed to take place near the subsolar point of the nucleus, which seems viable in that the activity of U3 is likely concentrated near the subsolar point (see Section \ref{subsec_morph}). If U3 is more dusty than we assumed, the required minimum active surface area can be even less.

Besides sublimation of supervolatiles, phase transition of the amorphous-crystalline water ice has been suggested to be a plausible mechanism for distant activity of a number of comets, as well as active Centaurs, at heliocentric distance $r_{\rm H} \approx 10$ au \citep[e.g.,][]{1992A&A...258L...9P,2009AJ....137.4296J}. Cometary nuclei are thought to be conglomerated from ices that were formed at 25 K \citep{2003Icar..162..183N,2005Icar..175..546N}. At such low temperatures, water ice condenses in an amorphous form because of lacking the energy to reorder into the crystalline lattice. Beyond $r_{\rm H} \gtrsim 10$ au, amorphous water ice can survive at the surface of cometary nuclei \citep{2012AJ....144...97G}. Upon heating, the phase transition from the amorphous to crystalline structure will occur, and process is exothermic and irreversible, during which formerly trapped gases in pores will be released \citep{1987PhRvB..36.9219L,1988PhRvB..38.7749B}. The crystallisation timescale is strongly dependent upon temperature:
\begin{equation}
\tau_{\rm c} = \tau_{\rm c, 0} \exp \left(-\frac{\mathscr{E}_{\rm A}}{k_{\rm B} T} \right)
\label{eq_tau_c},
\end{equation}
\noindent where $\tau_{\rm c,0} = 3.02 \times 10^{-21}$ yr is a scaling coefficient, $\mathscr{E}_{\rm A}$ is the activation energy, and $k_{\rm B}$ is the Boltzmann constant, with $-\mathscr{E}_{\rm A} / k_{\rm B} = 5370$ K \citep{1989ESASP.302...65S}. We equate the crystallisation timescale to the orbital period of U3 ($\sim$2 Myr) to determine the critical temperature $T_{\rm c}$ at which the onset of the phase transition occurs. This simplistic way only renders a highly conservative but nevertheless reasonable estimate, because we can be certain that no amorphous water ice will have survived if $\tau_{\rm c} \la 2$ Myr. Solving Equation (\ref{eq_tau_c}), we find $T_{\rm c} \approx 87$ K, which can be reached at $r_{\rm H} \approx 21$ au for inactive patches near the subsolar point of the nucleus, or $r_{\rm H} \approx 11$ au for the nucleus in the isothermal state. We can therefore conclude that by the time we started the first observation campaign at WIYN in late 2016, the onset of crystallisation of amorphous water ice may have already commenced, whereby trapped supervolatiles therein would be released. This coincided with the potential color variation of the comet, but as we mentioned, we do not know if they are related to each other.

Another proposed mass-loss mechanism for distant comets that is also related to the amorphous water ice is the annealing process, which can occur at temperature as low as $\sim$30 K \citep{1987PhRvB..35.2427B,2009Icar..201..719M}. However, we do not favor it for U3 because the mass flux of the gases \citep[see][Figures 2 \& 3]{2017MNRAS.469S.517N} is smaller than that due to sublimation of supervolatiles such as CO (or CO$_2$, in the subsolar case) by at least two orders of magnitude.

\begin{figure*}
\epsscale{1.0}
\begin{center}
\plotone{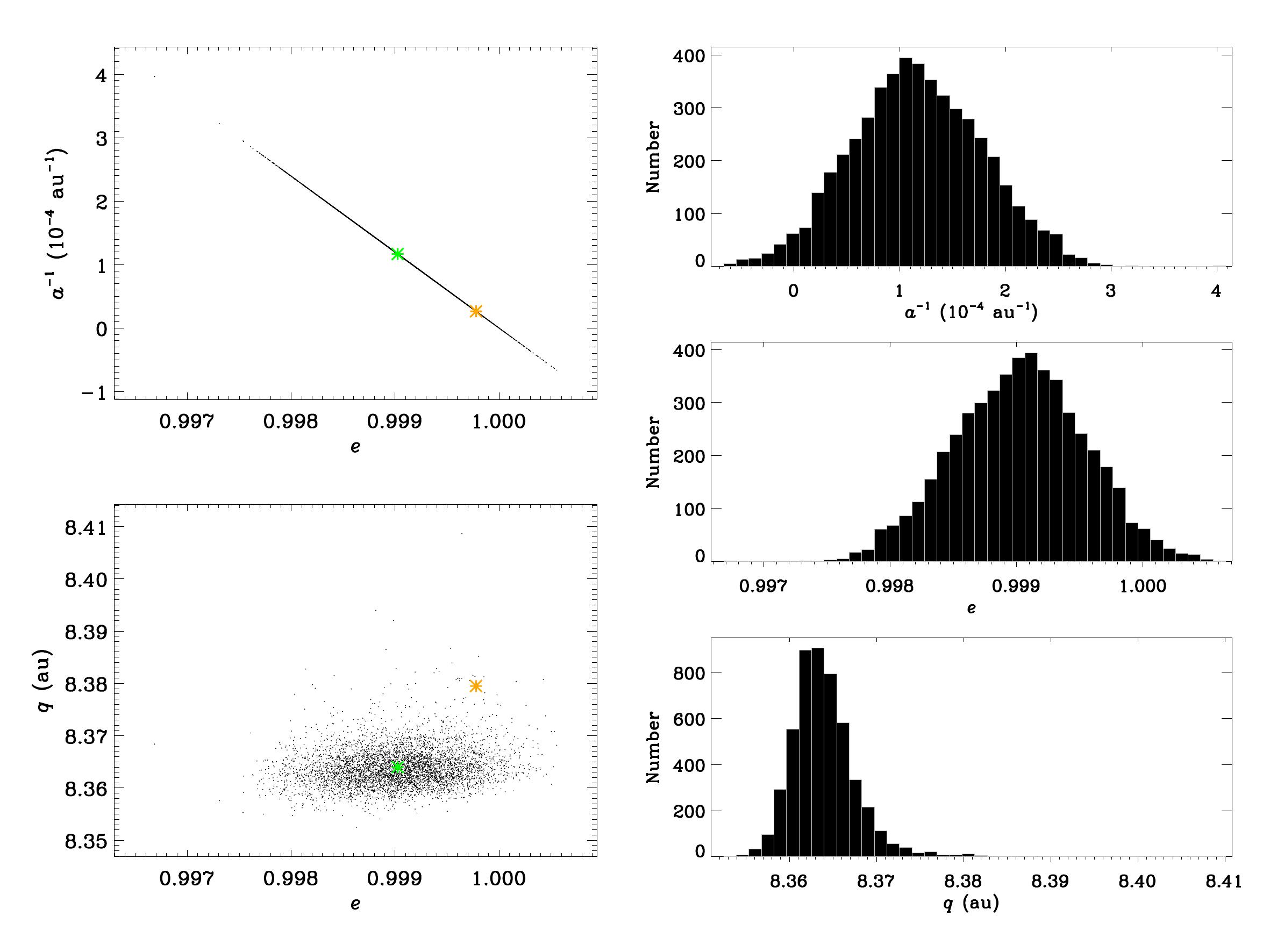}
\caption{
Distributions of the barycentric orbital elements of the 5001 clones of C/2010 U3 (Boattini) in the $a^{-1}$-$e$ and $q$-$e$ space at the previous perihelion (mean value $\left \langle t_{\rm p} \right \rangle = -1.96 \pm 0.04$ Myr from J2000). The nominal orbit and the mean orbit of the clones are marked as orange and green asterisks, respectively. Only a fraction of 2.6\% of the clones have barycentric $e \ge 1$.
\label{fig_U3_peri_prev}
} 
\end{center} 
\end{figure*}

\subsection{Past Dynamical Evolution}
\label{subsec_dynevo}

The ultra-distant activity of U3 draws our attention to its dynamical properties. To examine if the comet is visiting the planetary region for the very first time, i.e., dynamically new, we perform N-body integration analysis. Although the nongravitational effect  of the comet is not detected as expected, to encompass uncertainties thereof, we instead use the CO-sublimation $A_j$ ($j=1,2,3$) orbit solution (Table \ref{tab_orb}). Although this choice might allow larger nongravitational forces than the ones actually acting on U3, the advantage is that the uncertainty region in the propagated orbital element space in the gravity-only model will be fully engirdled by the one in the nongravitational model.\footnote{In fact, we have ran the same calculations with the gravity-only solution. Unsurprisingly, the results were found to be substantially the same, because the nongravitational forces, if any, are weak at great heliocentric distances.} We generate 5000 clones with initial conditions determined by the nominal orbit and the associated covariance matrix of the orbital elements from the CO-sublimation $A_j$ ($j=1,2,3$) model. Together with the nominal orbit, the clones are then backward integrated in our modified version of the {\it MERCURY6} package \citep{1999MNRAS.304..793C} with the gravity of the Sun, perturbations from the eight major planets and galactic tides \citep{2005CeMDA..93..229F}, and the relativistic corrections taken into consideration. Note that effects due to potential close encounters with nearby passing stars are excluded in our analysis, this is because the majority of the nearby stars in the neighbourhood of the solar system are low-mass dwarfs \citep{2001A&A...379..634G}, resulting in their perturbations to objects in the Oort cloud orders-of-magnitude smaller than that of the galactic tides. 

We obtain that, at epoch of -1 kyr from J2000, when the comet was at heliocentric distance $r_{\rm H} \approx 560$ au, the mean value of the barycentric reciprocal semimajor axis is $\left \langle a^{-1} \right \rangle = \left(6.37 \pm 0.08 \right) \times 10^{-5}$ au$^{-1}$. None of the clones have barycentric eccentricity $e \ge 1$ (to be more exact, $e < 0.9995$). Thus, we can confidently conclude that U3 is a comet from the Oort cloud. The backward integration is then continued until the previous perihelion is reached for all of the clones. The result of the statistics at the previous perihelion return is summarized and plotted in Figure \ref{fig_U3_peri_prev}. All of the clones have their previous perihelion passages in an epoch range of $-2.1 < t_{\rm p} < -1.8$ Myr from J2000, at barycentric perihelion distance $\left \langle q \right \rangle = 8.364 \pm 0.004$ au. In fact, 105 of the clones, including the nominal orbit, are found to have close approaches to Uranus within $\sim$1 au after passing the previous perihelion. Since we do not consider the orbital uncertainty of Uranus in our simulation, the complete geometry statistics of this potential encounter cannot be reliably established. Thus, we will not further interpret the encounter. Based upon that, during the previous perihelion, all of the clones have $r_{\rm H} < 8.5$ au, within the region where planetary perturbations are significant, we conclude that U3 is almost certain to be a dynamically old comet. However, the observed activity at great heliocentric distances cannot be accounted by any retained heat from the last apparition, because its thermal timescale is at least an order of magnitude shorter than the orbital period. It reinforces that there is possibly no clear correlation between dynamical history of a comet and its activity level, in support of the argument by \citet{2001A&A...375..643D}.

\subsection{Diversity in Ultra-Distant Comets?}
\label{subsec_diver}

Another ultra-distant comet we recently recognized is K2, which was observed to be active at $r_{\rm H} = 23.7$ au in prediscovery data \citep{2017ApJ...847L..19J,2017ApJ...849L...8M,2018AJ....155...25H}. Although K2 and U3 are both active at similar great heliocentric distance, their physical properties seem to be conspicuously distinguishable. While the former has been exhibiting a nearly circularly symmetric morphology since the discovery in mid 2017, which suggests ejection of submillimetre-sized or larger dust grains \citep{2018AJ....155...25H,2019AJ...157..65J}, the latter has been showing an obvious tail comprised of much smaller particles since the earliest prediscovery observations from 2005-2006. Not only are the two comets morphologically different, but they also exhibited dissimilar activity trends while approaching the Sun: while K2 has been almost steadily increasing its effective scattering cross-section \citep{2019AJ...157..65J}, U3 has shown fluctuations thereof indicative of instabilities. Based on our current knowledge, the most likely physical mechanism that can continuously drive their distant activity is sublimation of supervolatiles. Thus, our original expectation was that they would behave more or less alike if they are not compositionally distinct. However, perplexingly, this is not the case. Hitherto we only have several examples of ultra-distant comets. The existence of these differences seem to suggest a diversity amongst the population, which is possibly related to their birthplaces and evolutionary paths.

The hypothesis by \citet{2015A&A...583A..12G} that cohesion between particles at surfaces of a cometary nucleus probably provides an explanation for why particles smaller than submillimetre-sized have been held back at K2, but it fails for U3, which apparently manages to break this obstacle. In fact, even more disturbing is that, for K2 at heliocentric distance $r_{\rm H} > 10$ au, according to this hypothesis, the maximum ejectable grain sizes for overcoming the nucleus gravity are nonetheless smaller than the minimum ones for overcoming the cohesive forces between particles at the surface \citep{2019AJ...157..65J}. This means that one should not expect any cometary activity due to sublimation at that great heliocentric distance whatsoever, which clearly forms a contradiction to the actual observations of K2. This problem is even worse for U3, as the cohesion increases as the dust grain size shrinks at the nucleus surface.

Obviously, there is still a lot to be understood about how distant comets are active and whichever physical mechanism is at play driving the cometary mass loss. We thus strongly encourage future work on this subject.

\section{Summary}
\label{sec_sum}

We conclude our analysis of comet C/2010 U3 (Boattini) as follows:

\begin{enumerate}

\item The comet was observed to be active all the way back to 2005 November 05 at an inbound heliocentric distance of $r_{\rm H} = 25.8$ au, which is a new record. 

\item Despite the ultra-distant activity, we confidently identify the comet as a dynamically old member from the Oort cloud. The previous perihelion passage occurred at epoch $-1.96 \pm 0.04$ Myr from J2000, with barycentric $q = 8.364 \pm 0.004$ au.

\item The observed morphology of the comet are in great match with our Monte Carlo dust ejection models in which the gravitational force due to the Sun, Lorentz force and the solar radiation pressure force are considered altogether. Simulations without inclusion of the Lorentz force cannot match the observations.

\item Dust grains of $\sim$10 \micron~in radius are observed, which are ejected continuously at speeds of $\lesssim$50 m s$^{-1}$, consistent with sublimation of supervolatiles such as CO or CO$_2$. However, the observed activity of the comet at WIYN is also likely related to crystallisation of amorphous water ice, as this phase transition would have commenced at $r_{\rm H} \gtrsim 11$ au.

\item We find that the comet showed fluctuations in the effective scattering cross-section due to the two outburst events around 2009 and early 2017. Our estimated lower limit to the magnitude of the net mass-loss rate is only $\left| \dot{\mathcal{M}} \right| \sim 1$ kg s$^{-1}$. In order to suffice the activity, the nucleus radius is estimated to be $R_{\rm n} \gtrsim 0.1$ km, if CO (or other ices of similar volatility) is the dominant sublimating substance, or CO$_2$ from near subsolar points.

\item The general color of the comet is similar to those of other long-period comets, and redder than that of the Sun. Yet potential temporal variations are observed. In the {\it B} $-$ {\it V} wavelength interval, the comet reddened at $10 < r_{\rm H} < 15$ au, which coincided with crystallisation of amorphous water ice, if at all, but then gradually turned bluer and restored the original color. In the {\it V} $-$ {\it R} section, the comet likely reddened at $r_{\rm H} < 10$ au.

\item With the available astrometry and our refined measurements, a gravity-only orbital solution provides a satisfactory fit to the observational data; we cannot detect any significant nongravitational effects of the comet.

\end{enumerate}

\acknowledgements
{
We thank Bill Ryan for providing remeasured astrometry of the comet from the Magdalena Ridge Observatory, Melissa Brucker and Robert McMillan for providing information on the Spacewatch observations, and the anonymous referee for a prompt review. Discussions with David Jewitt and Yingdong Jia have greatly benefitted this research. The Keck data were provided by David Jewitt, who was the PI of the Keck observations. The WIYN images were taken with assists from Ariel Graykowski and Flynn Haase.
This research used the facilities of the Canadian Astronomy Data Centre operated by the National Research Council of Canada with the support of the Canadian Space Agency.
Based in part on observations at Kitt Peak National Observatory, National Optical Astronomy Observatory (NOAO Prop. IDs 2016B-0070, 2017A-0083, 2017B-0240, \& 2018B-0223; PIs: D. Jewitt, \& M.-T. Hui), which is operated by the Association of Universities for Research in Astronomy (AURA) under a cooperative agreement with the National Science Foundation. 
This research has made use of the Keck Observatory Archive (KOA), which is operated by the W. M. Keck Observatory and the NASA Exoplanet Science Institute (NExScI), under contract with the National Aeronautics and Space Administration.
DF conducted this research at the Jet Propulsion Laboratory, California Institute of Technology, under a contract with NASA.
}

\vspace{5mm}
\facilities{Keck(LRIS), WIYN:0.9m}

\software{IDL,
MERCURY6 \citep{1999MNRAS.304..793C}}

\end{document}